\documentclass[aps,pra,reprint,superscriptaddress]{revtex4-1}

\usepackage[pdftex]{graphicx}
\usepackage{upgreek}
\usepackage{epstopdf}
\usepackage{natbib}
\usepackage{color}
\usepackage [english]{babel}
\usepackage [autostyle, english = american]{csquotes}
\usepackage{hyperref,url}
\MakeOuterQuote{"}

\begin{document}


\title{STIRAP preparation of a coherent superposition of ThO $H^3\Delta_1$ states for an improved electron EDM measurement}


\author{C. D. Panda}
\email[]{cpanda@fas.harvard.edu}
\affiliation{Department of Physics, Harvard University, Cambridge, Massachusetts 02138, USA}

\author{B. R. O'Leary}
\affiliation{Department of Physics, Yale University, New Haven, Connecticut 06511, USA}

\author{A. D. West}
\affiliation{Department of Physics, Yale University, New Haven, Connecticut 06511, USA}

\author{J. Baron}
\affiliation{Department of Physics, Harvard University, Cambridge, Massachusetts 02138, USA}

\author{P. W. Hess}
\altaffiliation{Present address: Joint Quantum Institute, University of Maryland, College Park, Maryland 20742, USA}
\affiliation{Department of Physics, Harvard University, Cambridge, Massachusetts 02138, USA}

\author{C. Hoffman}
\altaffiliation[]{Present address: JILA, University of Colorado, Boulder, Colorado 80309, USA}
\affiliation{Department of Physics, Harvard University, Cambridge, Massachusetts 02138, USA}

\author{E. Kirilov}
\altaffiliation[]{Present address:: Institut f{\"u}r Experimentalphysik, Universit{\"a}t Innsbruck, A-6020 Innsbruck, Austria}
\affiliation{Department of Physics, Yale University, New Haven, Connecticut 06511, USA}

\author{C. B. Overstreet}
\altaffiliation[]{Present address: Department of Physics, Stanford University, Stanford, California 94305, USA}
\affiliation{Department of Physics, Harvard University, Cambridge, Massachusetts 02138, USA}

\author{E. P. West}
\affiliation{Department of Physics, Harvard University, Cambridge, Massachusetts 02138, USA}

\author{D. DeMille}
\affiliation{Department of Physics, Yale University, New Haven, Connecticut 06511, USA}

\author{J. M. Doyle}
\affiliation{Department of Physics, Harvard University, Cambridge, Massachusetts 02138, USA}

\author{G. Gabrielse}
\affiliation{Department of Physics, Harvard University, Cambridge, Massachusetts 02138, USA}



\begin{abstract}

Experimental searches for the electron electric dipole moment (EDM) probe new physics beyond the Standard Model. The current best EDM limit was set by the ACME Collaboration [Science \textbf{343}, 269 (2014)], constraining time reversal symmetry ($T$) violating physics at the TeV energy scale. ACME used optical pumping to prepare a coherent superposition of ThO $H^3\Delta_1$ states that have aligned electron spins. Spin precession due to the molecule's internal electric field was measured to extract the EDM. We report here on an improved method for preparing this spin-aligned state of the electron by using STIRAP. We demonstrate a transfer efficiency of $75\pm5\%$, representing a significant gain in signal for a next generation EDM experiment. We discuss the particularities of implementing STIRAP in systems such as ours, where molecular ensembles with large phase-space distributions are transfered via weak molecular transitions with limited laser power and limited optical access.
\end{abstract}

\date{\today}

\maketitle


\section{Introduction}

The current best upper limit on the electron electric dipole moment (EDM), $|d_e|<9.3\times10^{-29}~e \cdot \textrm{cm}$ (90$\%$ confidence), was set by the ACME Collaboration \cite{Baron2014} \footnote{Note that the limit reported here uses an updated value for $\mathcal{E}_{\rm eff}$ = 78 GV/cm, obtained from an unweighted mean of Refs. \cite{Skripnikov2015,Fleig2014}}. This represents an order of magnitude improvement on the previous limits on $d_e$ \cite{Hudson2011, Regan2002}. The ACME result significantly reduces the viable parameter space for time reversal symmetry ($T$) violating interactions between electrons and potential new particles at the TeV energy scale appearing in many extensions to the Standard Model \cite{Pospelov2005, Roberts2009}.

The ACME experiment (ACME I) performed a spin precession measurement \cite{Kirilov2013} on thorium monoxide (ThO) molecules from a cryogenic buffer gas beam source \cite{Hutzler2011}. The measurement took advantage of the high effective electric field $\mathcal{E}_{\rm eff}\approx 78$ GV/cm of the metastable ThO $H^3\Delta_1$ state, when the molecules are fully polarized in moderate electric fields ($\sim10$ V/cm) \cite{Skripnikov2015,Fleig2014,Meyer2008}. The use of the $H^3\Delta_1$ state allows for spectroscopic reversal of $\vec{\mathcal{E}}_{\rm eff}$ in the lab frame due to its $\Omega$-doublet structure \cite{Bickman2009}. Its small magnetic moment \cite{Petrov2014} greatly suppresses systematic errors related to magnetic fields and geometric phases \cite{Vutha2010}.

ACME I was limited by the 1-sigma statistical uncertainty in the EDM value of $\delta d_e \approx 1.5 \times 10^{-28}~e\cdot\textrm{cm} \sqrt{\textrm{day}}/ \sqrt{T}$, where $T$ is the running time in days (24 hours with realistic duty cycle) \cite{Baron2014}. The established systematic error limits were also limited by statistics, such that an improved statistical sensitivity is anticipated to result in an improved systematic uncertainty for an equal measurement time $T$. We are now modifying the ACME apparatus with the goal to improve the sensitivity of the EDM experiment.

In this paper, we discuss in detail a method that yields a significant increase in the number of useful ThO molecules for a next-generation ACME experiment (ACME II). This technique uses efficient optical transfer of population to prepare the initial spin-aligned state of the $H^3\Delta_1$ "EDM state." ACME I employed optical pumping to prepare this state, with an efficiency of approximately $6\%$. Using the technique of STImulated Raman Adiabatic Passage (STIRAP), we here demonstrate an increase in the population of the desired state by a factor of $12\pm1$, corresponding to a state transfer efficiency of $75\pm5\%$.

STIRAP is a population transfer scheme in a three-level system that relies on coherent two-photon coupling using time-varying electromagnetic fields \cite{Bergmann1998}. Under appropriate experimental conditions, STIRAP can nearly completely transfer population from an initially populated state $|1\rangle$ to a final state $|3\rangle$ via a possibly lossy intermediary state $|2\rangle$. The process relies on the adiabatic evolution of a "dark" (i.e. not coupled to the radiation fields), population-trapping state as molecules (or atoms) experience partially overlapping slowly varying fields: a Stokes pulse introduces a dynamic Stark splitting of the unpopulated states $|2\rangle$ and $|3\rangle$ that "clears the way" for a pump pulse, coupling states $|1\rangle$ and $|2\rangle$. After its discovery and first demonstration with Na$_2$ dimers \cite{Gaubatz1990}, STIRAP has been successfully applied to a number of experiments, such as in the preparation of ultracold dense gases of polar molecules \cite{Ni2008,Danzl2008}, creation of a well-defined photon number state in single-atom cavity quantum electrodynamics \cite{Parkins1993, Hennrich2000}, and quantum information processing \cite{Sorensen2006,Moller2007}.

STIRAP has been successfully implemented in such a multitude of systems by overcoming significant challenges, some generic to the method and others peculiar to the specific system. In experiments such as ours, where entire molecular ensembles with large phase-space distributions are transferred via weak molecular transitions, the required laser powers and intensities are significant and can be challenging to achieve with current laser technology. To perform STIRAP with near-unity transfer efficiency, two-photon resonance must be maintained \cite{Bergmann1998}, which can place demanding constraints on the phase coherence of the lasers used \cite{Yatsenko2014}. Furthermore, it is important to apply smoothly-varying Stokes and pump pulses in such a manner that the adiabaticity criterion is fulfilled during the entirety of the transfer process \cite{Kuklinski1989, Bergmann1998}. This is challenging due to the geometrical constraints of our apparatus.

\section{STIRAP Implementation}

\begin{figure} 
\includegraphics[scale=0.355]{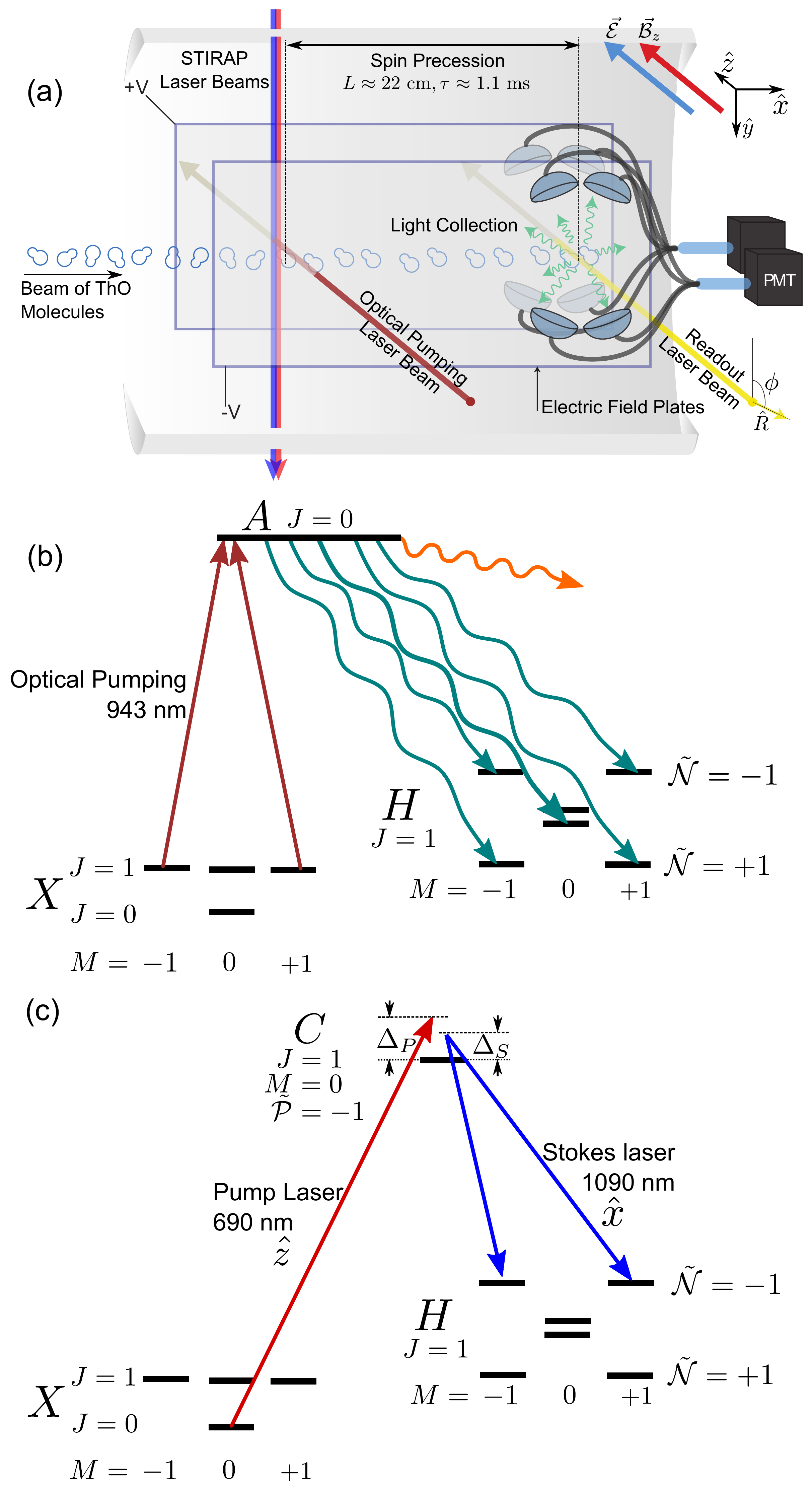}
\caption{(a) Schematic of the measurement apparatus used in the present work. A collimated pulsed beam of ThO molecules enters the interaction region. The spin-aligned state is prepared, precesses in parallel electric and magnetic fields and is read out in the detection region by linearly polarized light, with resulting fluorescence collected and detected by photomultiplier tubes (PMTs). The initial state can be populated using either (b) an optical pumping scheme similar to that used in ACME I, or (c) a STIRAP scheme. We alternate between the two methods for comparison purposes by blocking the corresponding laser beams. \label{Fig1}}
\end{figure}

The ACME experiment uses a pulsed beam of ThO generated by a cryogenic buffer gas beam source \cite{Hutzler2011}. The molecules exit the beam source with a forward velocity of $\sim$200 m/s along the $\hat{x}$ axis (Fig. \ref{Fig1}(a)). Their population is  primarily concentrated in the electronic ground state $X$, in several rotational states with a Maxwell-Boltzmann distribution corresponding to a rotational temperature of about 4~K. In the current measurements, the molecules are collimated 1.1 m downstream by a square aperture with dimensions of 25$\times$25 mm. The molecules exit the aperture with a "flat-top" transverse velocity distribution with full width at half maximum (FWHM) of about $4.5$ m/s. They travel into a region that has an uniform electric field $\vec{\mathcal{E}}$  that defines the $\hat{z}$ axis. The residual Earth's magnetic field has a component $\vec{\mathcal{B}}_z \sim 40$ mG along the $\hat{z}$ axis. The applied electric field magnitude $\mathcal{E}\approx75$ V/cm fully polarizes the molecule in the lab frame \cite{Vutha2010}. Thus, the internuclear axis, whose orientation $\hat{n}$ we define to point from the oxygen to thorium nucleus, is on average either aligned or anti-aligned with the applied electric field. This relative alignment is defined by the quantum number $\tilde{\mathcal{N}}\equiv \hat{\mathcal{E}}\cdot\hat{n}=\pm1$.


In ACME I, the efficiency of preparing the initial state was limited by the incoherent nature of the optical pumping process that was used (partially illustrated in Fig. \ref{Fig1}(b)) \cite{Baron2014}. There, the $|X,J=1,M=\pm 1,\tilde{\mathcal{P}}=-1\rangle$ states, where $\tilde{\mathcal{P}}$ is the parity quantum number, were optically pumped by two spatially separated 943 nm laser beams with orthogonal linear polarizations to the $|A, J=0,\tilde{\mathcal{P}}=+1\rangle$ state.  Approximately $35\%$ of the population excited by the laser subsequently spontaneously decayed into the $|H,J=1\rangle$ state manifold  \footnote{This efficiency number uses an updated calculation, which includes all paths of decay from the excited $A$ state, using data from the measurement discussed in Ref. \cite{Spaun2014} \label{note1}}. Decay to each of the mixed-parity states $|H,J=1,M=\pm1,\tilde{\mathcal{N}}\rangle$ occurred with equal probability ($P=1/6$), with decay to the odd parity state $|H,J=1,M=0,\tilde{\mathcal{P}}=-1\rangle$ being twice as likely  ($P=1/3$) and decay to the even parity state $|H,J=1,M=0,\tilde{\mathcal{P}}=+1\rangle$ being forbidden by the E1 parity selection rule. A linearly polarized 1090 nm laser beam, resonant with the $|H,J=1,M=\pm1,\tilde{\mathcal{N}}\rangle\rightarrow |C,J=1,M=0,\tilde{\mathcal{P}}\rangle$ transition, addressed the spectrally unresolved states  $|H,J=1,M=\pm1,\tilde{\mathcal{N}}\rangle$ of a particular $\tilde{\mathcal{N}}=\pm1$ quantum number, pumping out half of the population and leaving behind a "dark" coherent superposition. This coherent superposition corresponds to an electron spin-aligned state \cite{Kirilov2013}. For example, if the state preparation laser beam was linearly polarized along $\hat{x}$ and the $|C,J=1,M=0,\tilde{\mathcal{P}}=+1\rangle$ state was used, the prepared state was
\begin{equation}
|\psi(t=0),\tilde{\mathcal{N}}\rangle=\frac{ |M=+1,\tilde{\mathcal{N}}\rangle - |M=-1,\tilde{\mathcal{N}\rangle}}{\sqrt{2}},
\end{equation}
where  $|M=\pm1,\tilde{\mathcal{N}}\rangle$ is compact notation for $|H,J=1,M=\pm1,\tilde{\mathcal{N}}\rangle$.

The optical pumping transfer efficiency from the $|X,J=1,M=\pm1,\tilde{\mathcal{P}}=-1\rangle$  states to the $H$ state manifold is $\sim 35\%$  \cite{Note2}. One third of this population is contained in a pair of states with particular $\tilde{\mathcal{N}}=\pm1$ and half of the population is in the selected spin-aligned state. We therefore estimate the efficiency of transferring population from the $|X,J=1,M=\pm1,\tilde{\mathcal{P}}=-1\rangle$ states  to $|\psi(t=0),\tilde{\mathcal{N}}\rangle$ to be approximately 6$\%$ in ACME I.

In this work, we demonstrate the use of STIRAP to transfer population from the ro-vibrational ground state of the ThO molecule $|X,J=0,\tilde{\mathcal{P}}=+1\rangle$ directly into the desired spin-aligned state of the $H$ state with an efficiency of $75\pm5\%$ (Fig. \ref{Fig1}(c)).

In ACME I, the population in the $|X,J=1,M=\pm1,\tilde{\mathcal{P}}=-1\rangle$ states was enhanced with population from the other rotational levels through optical pumping and microwave transfer by a factor of about 1.5--2.0. Alternate rotational cooling schemes can provide roughly the same population in the $|X,J=0,\tilde{\mathcal{P}}=+1\rangle$ state as was previously available in both $|X,J=1,M=\pm1,\tilde{\mathcal{P}}=-1\rangle$ states combined. The gain in usable molecules in a future EDM measurement can then be parameterized as $G=g_{\rm RC}\cdot g_{\rm ST}$, with rotational cooling gain $g_{\rm RC}\approx 1$. From here on, we refer only to the STIRAP improvement factor $g_{\rm ST}$.

The $|X,J=0,\tilde{\mathcal{P}}=+1\rangle \equiv |1\rangle $ initial state and desired spin-aligned state $|\psi(t=0),\tilde{\mathcal{N}}\rangle \equiv |3\rangle$ (from Eq. 1) are one unit of angular momentum projection apart ($\Delta M=\pm1$). STIRAP between $|1\rangle$ and $|3\rangle$ requires one laser beam to have $\hat{z}$ polarization (corresponding to $\Delta M=0$) and one $\hat{x}$ polarization (corresponding to $\Delta M=\pm 1$). Access to the $\hat{z}$ polarization in the molecular beam region requires that the laser beams be sent vertically (along the $\hat{y}$ axis), as shown in Fig. \ref{Fig1}(a). In this configuration, the laser fields do not transmit through the transparent field plates, as was the case in ACME I. This avoids potential optical damage to the electric field plates and prevents imperfect STIRAP laser intensity profiles due to interactions (e.g. reflections) with the field plates. 

The radiative couplings between the three levels are characterized by the Rabi frequencies
\begin{equation}
\Omega_i(t)=\frac{\vec{D}_i \vec{E}_i(t)}{\hbar},
\end{equation}
where $i \in \{\rm{S}, \rm{P}\}$ corresponds to the Stokes or pump transition, $\vec{D}_i$ is the transition dipole moment, and $\vec{E}_i= E_i \hat{\epsilon}_i$ is the vector amplitude of the laser radiation field, which includes polarization. As shown in Figure \ref{Fig1}(c), the STIRAP efficiency is usually parametrized as a function of the detunings of the pump and Stokes lasers from their respective one-photon resonances $\Delta_P$ and $\Delta_S$. The STIRAP transfer efficiency is significantly more sensitive to the two-photon detuning $\delta=(\Delta_P-\Delta_S)/2$  than the one-photon detuning $\Delta=(\Delta_P+\Delta_S)/2$ \cite{Bergmann1998}.

As shown in Fig. \ref{Fig1}(c), we perform STIRAP via couplings to the short-lived $|C,J=1,M=0,\tilde{\mathcal{P}}=-1\rangle \equiv |2\rangle$ state, which has a lifetime of $\tau_C=500$ ns \cite{Hess2014}. The pump laser beam (690 nm) is linearly polarized along $\hat{z}$ and is near-resonant with the transition between states $|1\rangle$ and $|2\rangle$. The Stokes laser (1090 nm) is linearly polarized along $\hat{x}$ and is near-resonant with the transition between $|2\rangle$ and $|H,J=1,M=\pm1,\tilde{\mathcal{N}}\rangle$. Due to the parity of the intermediary state, population is transferred into the spin-aligned state $|3\rangle$ given in Eq. 1 \cite{Kirilov2013}. In the ACME I optical pumping scheme, the orthogonal spin orientation could be prepared by choosing the opposite excited state parity, $|C,J = 1,M = 0,\tilde{\mathcal{P}}\rangle$, or by rotating the polarization of the depletion 1090 nm laser to the $\hat{y}$-direction to address the opposite spin superposition. This was used as a "switch" for rejection of systematics \cite{Baron2014}. In this STIRAP scheme, the orthogonal spin orientation cannot be prepared, but we believe that this will not significantly impact our systematic error.

After being prepared by either STIRAP or the ACME I method, electric and magnetic fields cause the spin-aligned state to accumulate a phase $\phi$, resulting in
\begin{equation}
|\psi(\tau),\tilde{\mathcal{N}}\rangle=\frac{e^{-i \phi}|M=+1,\tilde{\mathcal{N}}\rangle-e^{+i\phi}|M=-1,\tilde{\mathcal{N}}\rangle}{\sqrt{2}}.
\end{equation}
The phase $\phi$ is dominated by the effects of $|\mathcal{B}_z|=|\vec{\mathcal{B}} \cdot \hat{z}|$, and its $\tilde{\mathcal{B}}=\textrm{sgn}(\vec{\mathcal{B}} \cdot {\hat{z}})$
\begin{equation}
\phi \approx \frac{-\mu_{\rm B} g \tilde{\mathcal{B}} |\mathcal{B}_z|\tau}{\hbar},
\end{equation}
where $g$ is the g-factor and $\mu_{\rm B}$ is the Bohr magneton. This phase also describes the angle by which the initial spin alignment rotates in the x-y plane while in the interaction region.

After traveling through the interaction region for a distance $L \approx 22$ cm, corresponding to a time $\tau \approx$ 1.1 ms, the phase of the spin-aligned state can be read out in the detection region using laser-induced fluorescence. The detection scheme relies on excitation of molecules with a linearly polarized readout laser to a short-lived state (the $C$ state in ACME I, the $I$ state in the current work) that emits photons when decaying to the ground state $X$. These photons are collected and detected by PMTs. In ACME I, $\phi$ was determined by rapidly switching between polarizations. In the STIRAP  experiments described here, we are primarily interested only in the total number of molecules, so the laser polarization direction is not switched, but rather kept constant. The laser's linear polarization direction $\hat{R}$ is chosen such that the detected fluorescence signal is maximized.

Care must be taken to minimize contamination of the detected signal with photons from other regions of the experiment, which would lead to additional noise in our EDM data. The STIRAP scheme we describe in this paper uses pump coupling through the same $X\rightarrow C$ transition that was used in ACME I for fluorescence detection (690 nm). We avoid the background from the STIRAP pump laser by using excitation at 703 nm from the $H$ state to the $I$ state instead of the $C$ state \cite{Kokkin2014}. The fluorescence accompanying the $I$ state decay at 512 nm is easily separable from the 690 nm pump light background. The $I$ state has all of the necessary features for the future ACME II detection scheme: a large branching ratio to $X$, a small branching ratio to $H$, a strong enough transition to $H$, and spectroscopically resolved states of opposite parity \cite{Kokkin2015,Edvinsson1968}.

\subsection{Laser system}

The STIRAP transfer is implemented with light derived from two systems of commercial external cavity diode lasers (ECDLs) at 1090 nm (Stokes laser, near-resonant with the  $C\rightarrow H$ transition) and 690 nm (pump laser, near-resonant with the $X\rightarrow C$ transition) \footnote{All four lasers are Toptica DL pro with antireflection (AR)-coated diodes}. The lasers are actively frequency stabilized through simultaneous locking to a horizontal cylindrical ultra low expansion (ULE) glass cavity with a finesse of 30,000. The cavity is housed in a lab-built evacuated aluminum enclosure with two stages of temperature control. It is temperature regulated near the critical temperature of the ULE spacer of 27.8$^\circ$C with a long-term (usually days) stability better than 1 mK. 

The lasers are locked to the ULE resonator using a feedback system based on the Pound-Drever-Hall (PDH) locking scheme \cite{Drever1983}. The feedback is provided by a commercial digital proportional-integral-derivative (PID) regulator, with a fast component to the current (bandwidth up to 5 MHz), and a slow component to the grating piezo (bandwidth up to 100 Hz).  A resonant EOM at 10 MHz creates the modulation sidebands necessary for the PDH lock, and an AOM allows for freely tuning the laser frequency away from the resonances of the cavity. A commercial fiber frequency comb locked to a GPS-stabilized RF reference allows for monitoring and long term correction of the absolute laser frequencies. The largest correction is a linear drift of $\sim$7 kHz/day, presumed to be due to the mechanical relaxation of the ULE spacer.

The 690 nm laser light is amplified by a commercial tapered amplifier (TA) to $\sim$300 mW. The output of the TA is then coupled through a single-mode fiber that also acts as a spatial filter for light delivered to the experiment. The 1090 nm laser is amplified by a commercial fiber amplifier to $\sim$10 W. 

The readout laser light at 703 nm is produced by a commercial Ti:Sapphire laser \footnote{Commercial MSquared Solstis 4000 system.}. Its frequency is actively stabilized through an offset beat note lock with an 703 nm ECDL \cite{Schunemann1999}, which is locked to the ULE cavity in the same manner as described above. Feedback is applied to the fast and slow piezo that positions one of the mirrors of the Ti:Sapphire cavity, with a bandwidth of up to 100 kHz. The linewidths observed were approximately 20 kHz (100 ms integration time), with a long term stability of 100 kHz within 24 hours.

\subsection{Phase noise}

Population transfer with close to unity efficiency of the entire ensemble of ThO molecules is possible in STIRAP if the two-photon detuning of the laser fields is near zero, i.e. roughly within the two-photon population transfer linewidth $\Delta \omega_{\rm 2ph}$ \cite{Romanenko1997}. In the case of significant differential phase noise between the pump and the Stokes lasers outside of this two-photon resonance linewidth, but within the one-photon resonance linewidth, the dark state eigenvector can acquire a component of the intermediary state $|2\rangle$ and population can radiatively decay out of the three-level system. The two-photon lineshape is difficult to describe analytically and varies significantly with the specific properties of the system, such as molecule phase space distribution, Rabi frequencies, lifetime of the excited state, interaction time, and one-photon detuning. Nevertheless, in our system it is possible to crudely estimate $\Delta \omega_{\rm 2ph}$ as follows.  Due to experimental constraints described below, we operate in a regime where the time when the STIRAP pulses overlap $\Delta T$ is on the order of $1/\gamma_C$, where $\gamma_C$ is the decay rate of the intermediate excited state $C$. In this case, $\Delta \omega_{\rm 2ph}$ is within a factor of order unity of $\Omega_{\rm eff}/2$, where $\Omega_{\rm eff}=\sqrt{\Omega_P^2+\Omega_S^2}$ is an effective two-photon Rabi frequency \cite{Romanenko1997}. In our system, two-photon linewidths $\Delta \omega_{\rm 2ph}$ are typically in the range of $2\pi \times (2$--$4) $ MHz FWHM, and $\Omega_{\rm eff}\approx2\pi \times 14$ MHz,  limited by the pump and Stokes laser beam intensities.

\begin{figure} 
\includegraphics[scale=0.67]{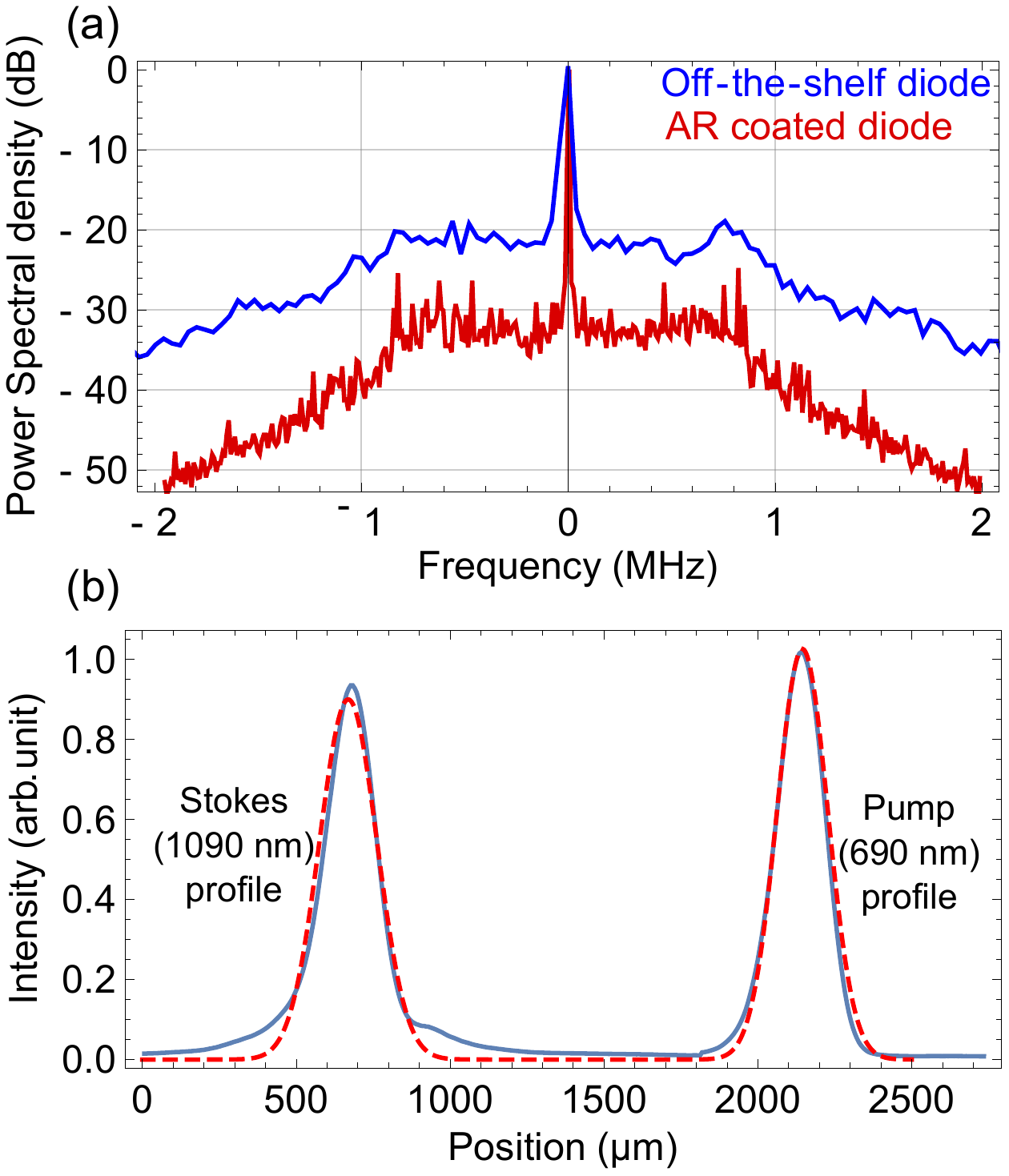}
\caption{(a) Power spectral densities of the 690 nm lasers are typical of stabilized ECDLs, with a very narrow central component on top of a much weaker broad pedestal. The AR-coated diodes provide a 10 dB reduction in the noise pedestal compared to the off-the-shelf diodes. (b) Typical laser beam intensity profiles (along the $\hat{x}$ axis) at the center of the molecular beam (integrated over a 5 mm range along $\hat{z}$ - blue line)  and Gaussian fits (dashed red line). The beam profiles are greatly separated for clarity. \label{Fig2}}
\end{figure}

To characterize the phase noise present in our lasers, we measured their power spectral densities using an optical beat note with a second laser system locked to a separate ULE cavity. Both systems were constructed to be as similar as possible. Figure \ref{Fig2}(a) shows the power spectral density of a beat note between two identical 690 nm lasers (50 ms integration time). The power spectra show a narrow $\approx 2\pi \times$150 Hz FWHM peak on top of a much broader suppressed pedestal, typical of stabilized ECDLs. The width of the central peak was measured using a much smaller resolution bandwidth than shown in Fig. \ref{Fig2}(a). The first tests were performed with a set of off-the-shelf diodes (blue in Fig. \ref{Fig2}(a)), which showed relatively high power in the spectral pedestal. Using these diodes, we observed  that the STIRAP transfer efficiency saturated as a function of laser power, reaching a maximum level of only 30--40$\%$. This behavior is consistent with recently reported detrimental effects on STIRAP transfer efficiency due to broad pedestals in the spectral lineshape that are common to stabilized ECDLs \cite{Yatsenko2014}. Furthermore, we performed simulations of the STIRAP efficiency by integrating the Lindblad master equation with random sampling over the spatial and velocity distributions within the molecular beam. The simulations were run with a phase model \footnote{The power spectral density of the phase follows a Lorentzian distribution. In the limit that the integral of the phase power spectral density is small compared to 1, the laser field spectrum consists approximately of a delta function at the center frequency that contains most of the power and a Lorentzian pedestal elsewhere.} consistent with the measured power spectral density of the lasers, varying phase noise parameters such as pedestal amplitude, width and position of noise features within the uncertainty of the measurement. The results of these simulations are consistent with the observed behavior, i.e. they showed a typical STIRAP saturation with laser power of the transfer efficiency at only 30--40$\%$ in a parameter range similar to that of our system. 

To reduce the laser phase noise in the pedestals, the off-the-shelf diodes were replaced with AR-coated diodes, which yield a narrower linewidth in an ECDL configuration before locking ($2\pi \times 200$ kHz compared to $2\pi \times1.5$ MHz). The power spectrum of the beat note from locked ECDLs with the new diodes displays a much-suppressed pedestal ($-30$ dB instead of $-20$ dB) with approximately the same pedestal linewidth ($\sim2\pi \times 2$ MHz FWHM), as shown in Fig. \ref{Fig2}(a). Simulations using a phase noise model consistent with the improved power spectrum predict near-unity efficiency. Due to changes to the experiment geometry that occurred at the same time as the reduction in laser phase noise, we could not verify empirically that the phase noise was directly responsible for the previously observed low transfer efficiencies. The 1090 nm laser power spectral densities, not shown here, exhibit similar pedestal suppression and pedestal linewidths to those of the 690 nm AR-coated diodes.

\subsection{ Particularities of the ACME experiment}

Although STIRAP has been performed in a number of atomic and molecular beam experiments \cite{Gaubatz1990,Ni2008,Danzl2008,Parkins1993,Hennrich2000,Sorensen2006}, each system presents its own difficulties. STIRAP within the ACME experiment is challenging for several reasons, which mostly arise from the fact that we are operating with lasers with power outputs that are close to our minimum requirements for efficient population transfer.

In considering the spatial intensity profiles necessary for the laser beams, it is important to note that STIRAP relies on adiabaticity for obtaining high transfer efficiencies \cite{Bergmann1998}. The "local" adiabaticity criterion,
\begin{equation}
\left|\frac{\dot \Omega_P \Omega_S-\Omega_P \dot \Omega_S}{\Omega_P^2+\Omega_S^2} \right| \ll |\omega^{\pm}-\omega^0|,
\end{equation}
where $|\omega^{\pm}-\omega^0|$ is the field-induced splitting in the dressed state energy eigenvalues \cite{Kuklinski1989}, sets constraints on the spatial "smoothness" and overlap of the STIRAP laser beams. In the case when the laser profiles have smooth shapes, an integration of the above gives the "global" adiabaticity criterion $\Omega_{\rm eff} \Delta T \gg1$ \cite{Bergmann1998}.

Laser beamshaping is restricted by optical power availability and geometrical considerations: in the $\hat{z}$ direction, the laser beams need to be significantly larger than the 25 mm diameter of the molecular beam in order to ensure that all molecules are addressed. The laser beam diameters necessary along the molecular beam forward velocity (along $\hat{x}$) are constrained by both the adiabaticity and two-photon resonance conditions. It can be shown that $\Omega_{\rm eff} \Delta T \propto \sqrt{w_x}$, where $w_x$ is the waist (1/$e^2$ intensity half-width) of the laser beam along $\hat{x}$ and the proportionality constant is a function of the transition dipole moments, available laser power, laser beam diameter along $\hat{z}$, and molecular beam longitudinal velocity. The adiabaticity criterion, $\Omega_{\rm eff} \Delta T \gg1$, puts a lower limit on the waist at $w_x \gg 10$ $\mu$m. 

The transverse velocity distribution of the ThO molecules in the STIRAP transfer region has a FWHM of 4.5 m/s, as discussed in the introduction to Section II. Since the two photons have a large relative wavelength difference (690 nm pump and 1090 nm Stokes), the different velocity classes will experience different Doppler shifts for the two STIRAP beams. This results in a distribution of two-photon detunings within the molecular beam with a width of $\Delta \omega_{\rm Doppler}\approx2\pi \times 1.2$ MHz FWHM. $\Delta \omega_{\rm Doppler}$ must be smaller than the intrinsic two-photon linewidth of the STIRAP process $\Delta \omega_{\rm 2ph}$ to ensure near-unity transfer efficiency for all of the molecules in the ensemble. $\Delta \omega_{\rm 2ph}$ increases with increasing laser intensity \cite{Halfmann2003}, which we achieve by decreasing the laser beam waists along the molecular beam forward velocity, $w_x$. Simulations involving numerical integration of the Hamiltonian for the three-level system show that for close to unity transfer efficiency, we require $w_x < 300$ $\mu$m.

These limits are also set, in part, by the weak transition dipole moment of the Stokes transition. The $H\rightarrow C$ dipole moment is estimated at $0.02$ $ea_{0}$ \cite{Spaun2014}. Even with 10 W of power available for the Stokes transition, the typical peak Rabi frequencies accessed in our system, $\Omega_S\approx 2 \pi \times$8 MHz, are orders of magnitude smaller than available in the first demonstration of STIRAP \cite{Gaubatz1990}. For comparison, the pump transition X$\rightarrow$ C dipole moment is estimated at $0.3$ $ea_0$ \cite{Spaun2014,Hess2014}, making the power requirements lower for that transition. With 50 mW of power, we are able to achieve pump Rabi frequencies of $\Omega_P\approx 2\pi \times 12$ MHz.

To satisfy previously described constraints, the laser beams are expanded in $\hat{z}$ to waists of 20--25 mm and then collimated. Along $\hat{x}$, the optical beams are first expanded to diameters of 10--20 mm and then focused to the required small waists ($w_x$) of $150$ $\mu$m (690 nm pump) and  $160$ $\mu$m (1090 nm Stokes), at the position of the molecular beam. The Rayleigh lengths for the laser beams are $100$ mm (pump) and $70$ mm (Stokes), larger than the molecular beam diameter of 25 mm, ensuring small variations in the laser beam diameter and peak intensity across the molecular beam along the vertical direction $\hat{y}$. The resulting peak intensities are $I_S\approx 1000$ mW/mm$^2$ for the Stokes beam (1090 nm) and $I_P\approx 6$ mW/mm$^2$ for the pump beam (690 nm).

Due to experimental complexities associated with other components of the ACME experimental apparatus, such as large-volume mu-metal magnetic shields, experiment vacuum chamber, and magnetic field coils, the available optical access is limited. To allow for easy adjustability, the last optical element is placed outside of the magnetic shields, at a distance of 1.5 meters from the focal point. These constraints limit the achieved laser intensity profile quality. Figure \ref{Fig2}(b) shows profiles of the laser beams along the molecules' forward velocity axis ($\hat{x}$), at the waist, measured with a CCD beam profiler. It is important to note that the quality of the beam shapes degrades at the vertical extremities of the molecular beam, as one moves away from the focal point, with Airy-like lobes in the tails increasing in amplitude up to 10--20$\%$ of the maximum intensity. 

Imperfect laser intensity profiles can either cause the local adiabaticity criterion (Eq. 5) to not be fulfilled, leading to non-adiabatic transfer of population to the intermediary lossy state $|2\rangle$, from which it decays out of the system, or can leave population in the initial state $|1\rangle$. Additionally, when on one-photon resonance, excess optical power in wings of the laser beam profiles similar to an Airy pattern caused by clipping of the laser beams can drive optical pumping, depleting the population of the initial state $|1\rangle$ before the STIRAP process begins, or depleting the desired final state $|3\rangle$ after the two-photon process is complete. Careful alignment of the relative pointing of the Stokes and pump beams to better than a few milliradians is extremely important for maintaining optimal overlap over the vertical 25 mm spatial extent of the molecular beam.

\subsection{Gain measurement} 

In order to quantify the signal improvement over ACME I, we measure the STIRAP molecule gain factor $g_{\rm ST}$ by quickly switching between the STIRAP state preparation scheme and the ACME I optical pumping scheme by alternately blocking the relevant laser beams for these two schemes on a timescale of 5 seconds, faster than normal fluctuations in the molecule beam flux \cite{Hutzler2011}. We detect the population in the prepared spin-aligned state by optically pumping on the $H\rightarrow I$ transition with a linearly polarized laser and detecting laser induced fluorescence signals at 512 nm ($S_{\rm ST}$ and $S_{\rm OP}$, respectively, for STIRAP and optical pumping) that are proportional to the transferred population.

Given that STIRAP state transfer is performed out of $|X,J=0,\tilde{\mathcal{P}}=+1\rangle$, and the ACME I state transfer scheme was performed out of $|X, J=1, M=\pm1,\tilde{\mathcal{P}}=-1\rangle$, the gain factor $g_{\rm ST}$ can be expressed as
\begin{equation}
g_{\rm ST} = \frac{S_{\rm ST}}{S_{\rm OP}} \frac{\sum_{\pm}P(|X,J=1,M=\pm1,\tilde{\mathcal{P}}=-1\rangle)}{P(|X,J=0,\tilde{\mathcal{P}}=+1\rangle)},
\end{equation}
where $P(|\psi\rangle)$ is the initial population in state $|\psi\rangle$. In the following measurements, rotational cooling schemes are not used. The initial populations are assumed to follow a Maxwell-Boltzmann distribution with a rotational temperature of $4\pm1$~K \footnote{A Neon buffer gas flow of 40 SCCM was used in these measurements.}, as observed previously with our beam source \cite{Hutzler2011}.

\section{Experimental Results}

The STIRAP transfer efficiency $\eta_{\rm ST}$ is calculated from the measured gain by $\eta_{\rm ST}=k \cdot g_{\rm ST}$, where the proportionality factor $k$ is obtained from an auxiliary calibration measurement. This calibration was performed by measuring both gain and transfer efficiency under experimental conditions where efficiency can be extracted with ease. We performed STIRAP with a one-photon detuning of $\Delta=2\pi \times 15$ MHz, much larger than the Doppler linewidths of the pump (690 nm) and Stokes (1090 nm)  beams of $2\pi \times 3.2$ MHz HWHM and $2\pi \times 2$ MHz HWHM respectively. The detuning is also much larger than the one-photon natural linewidth of $2\pi \times 0.3$ MHz HWHM. This large one-photon detuning is chosen such that one-photon transitions are highly suppressed. Even in the case of non-adiabatic transfer, the population that is not transferred to the desired final state $|3\rangle$ remains in the initial state $|1\rangle$ rather than populating the lossy intermediary state $|2\rangle$ and decaying out of the three-level system. A second probe laser, at 690 nm, driving the $|1\rangle \rightarrow |2\rangle$ transition, produces laser-induced fluorescence proportional to the population in state $|1\rangle$. The resulting 690 nm fluorescence is detected with the same PMTs used to detect the 512 nm fluorescence from the $I$ state, as discussed above. 

Fluorescence signals proportional to the population in state $|1\rangle$ are recorded when performing STIRAP ($S^{|1\rangle}_{\rm ST}$) and normalized to the case when no excitation is present ($S^{|1\rangle}_{0}$). The fluorescence signal proportional to leftover population in state $|1\rangle$ after performing STIRAP is given by
\begin{equation}
S^{|1\rangle}_{\rm ST}=S^{|1\rangle}_{0}[1-\eta_{\rm ST} (1-\eta_{\rm decay})], \label{eff}
\end{equation}
where the correction factor $(1-\eta_{\rm decay}$) accounts for decay from the metastable state $H$  back to state $|1\rangle$ during the time molecules travel between the population transfer and the readout regions. Vibrational and rotational branching ratios \footnote{A vibrational branching from ($H,v=0$) to ($X,v=0$) of $\sim 94\%$ is calculated using the method of Ref. \cite{Nicholls1981} and spectroscopic data from Ref. \cite{Edvinsson1985}. Rotational branching from $(H,J=1,M=\pm1 )$ to $(X,J=0)$ is $1/3$.} along with the lifetime of the $H$ state ($\sim$ 2 ms) \cite{Baron2014} and the precession time ($\tau \approx$ 1.1 ms) give an estimated $\eta_{\rm decay}\approx 13\%$, which is the fraction of population in the $H$ state that decays back to state $|1\rangle$. Under the same large one-photon detuning conditions corresponding to the calibration measurement,  $\eta_{\rm ST}$ is extracted from Eq. \ref{eff}, and the gain $g_{\rm ST}$ is measured using the same procedure described above. We then calculate the proportionality factor $k=\eta_{\rm ST}/g_{\rm ST}\simeq6.2\pm0.3$, where the uncertainty is dominated by the error in the rotational Boltzmann factor. We then use $k$ to infer the transfer efficiency $\eta_{\rm ST}$ from the measured gain $g_{\rm ST}$ for all other data, regardless of the laser detunings and other experimental parameters.

\begin{figure} 
\includegraphics[scale=0.45]{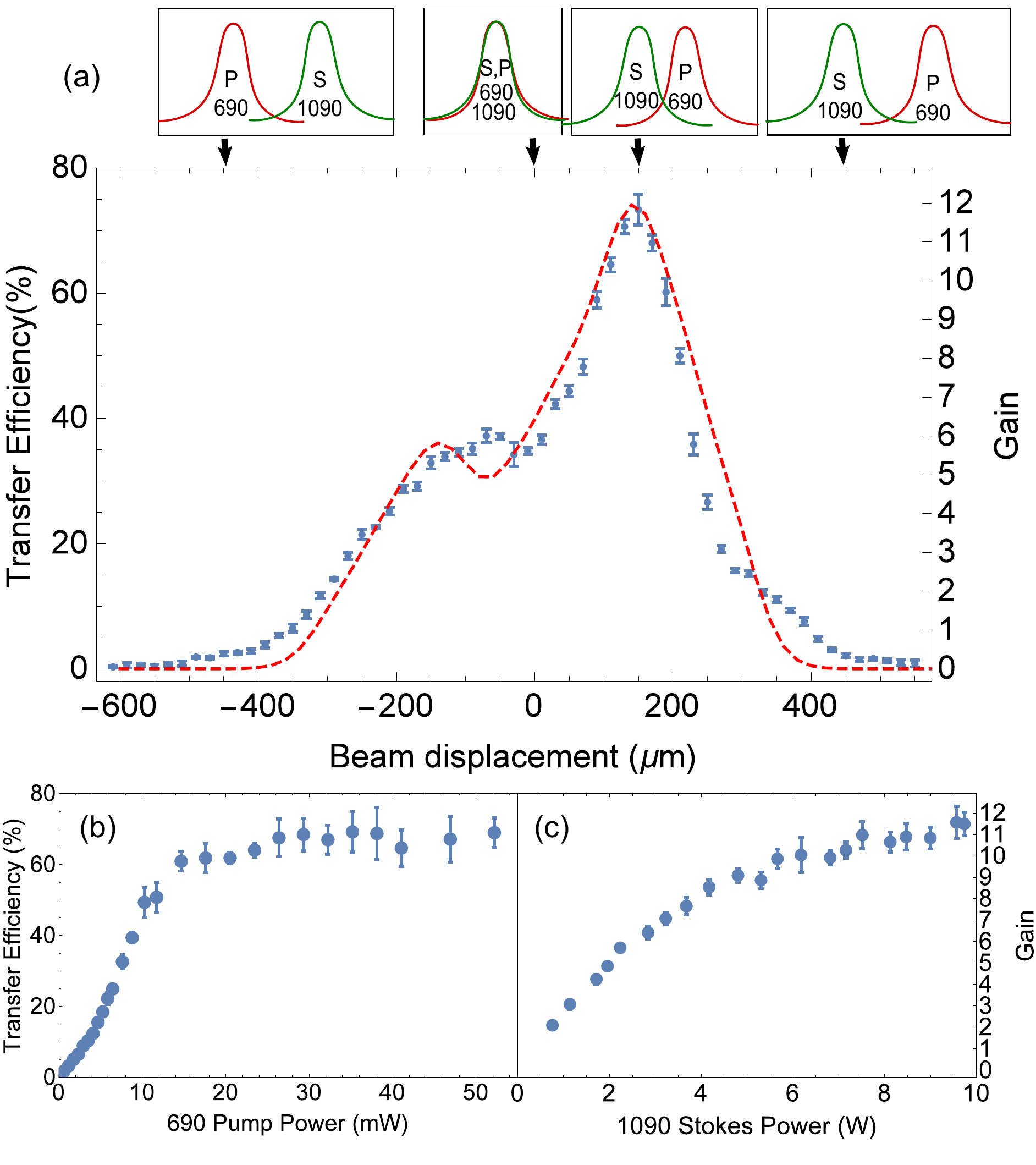}
\caption{(a) Efficiency of population transfer from ground state $|X,J=0\rangle$ to a spin-aligned state in the $|H,J=1\rangle$ state, as a function of the spatial overlap of the pump and Stokes beams (laser beam widths $\approx$ 150 $\mu$m) for the experiment (blue) and simulation (red dashed line). Insets along the top show the relative positions of the optical field pulses, as they are encountered by the molecules, traveling from left to right. Power saturation behavior of the (b) pump 690 nm laser and (c) Stokes 1090 nm laser. \label{Fig3}}
\end{figure}

The transfer efficiency measured under optimal conditions in our system is shown in Fig. \ref{Fig3}(a), as the spatial overlap of the pump and Stokes beams is varied, at a one-photon detuning $\Delta= 2\pi \times 8$ MHz. The two-photon resonance condition, $\delta=0$, is maintained for molecules with zero transverse velocity. As one would expect from the underlying theory \cite{Bergmann1998}, optimal transfer efficiency is obtained for the Stokes pulse preceding the pump pulse, with a separation between the two comparable to the waist size (160 $\mu$m). The maximum observed transfer efficiency is $75\pm5 \%$.  We observe a second, lower efficiency, local maximum when the two laser beams overlap in the reverse order, with the pump pulse applied first. This feature is a consequence of a large one-photon detuning \cite{Gaubatz1990}. Unlike when on resonance, the initial state is not optically pumped by the pump pulse arriving first. As the molecules pass through the laser beams, the overlap region allows a partially adiabatic two-photon process to drive a fraction of the population ($40\%$) to the final state $|3\rangle$. We observe a dip in efficiency when the two pulses are completely overlapped and the transfer efficiency vanishes when the separation between the laser beams is large compared to the laser intensity widths, as the overlap between the Stokes and pump beams drops to zero. Simulations performed by integrating the Schroedinger equation with a three-level system Hamiltonian show qualitative agreement to the data (Fig. \ref{Fig3}(a)). The simulations were performed with molecular ensemble parameters consistent with the measured experimental values. In addition, a Stokes and pump relative laser beam pointing misalignment of $\sim$5 mrad was included in the simulation and partially accounts for the lower than unity efficiency.

Figures \ref{Fig3}(b) and \ref{Fig3}(c) show the dependence of transfer efficiency on the power of the Stokes and pump lasers. Each data set is taken with the other laser at full power, always at the same one-photon detuning $\Delta=2\pi \times 8$ MHz and on two-photon resonance $\delta=0$, for molecules with zero transverse velocity. The 690 nm pump transition $|1\rangle \rightarrow |2\rangle$ is driven well into the saturated regime. The STIRAP transfer efficiency vs. 1090 nm Stokes laser power data shows that higher transfer efficiency might be achievable with greater power (Fig. \ref{Fig3}(c)).

\begin{figure} 
\includegraphics[scale=0.77]{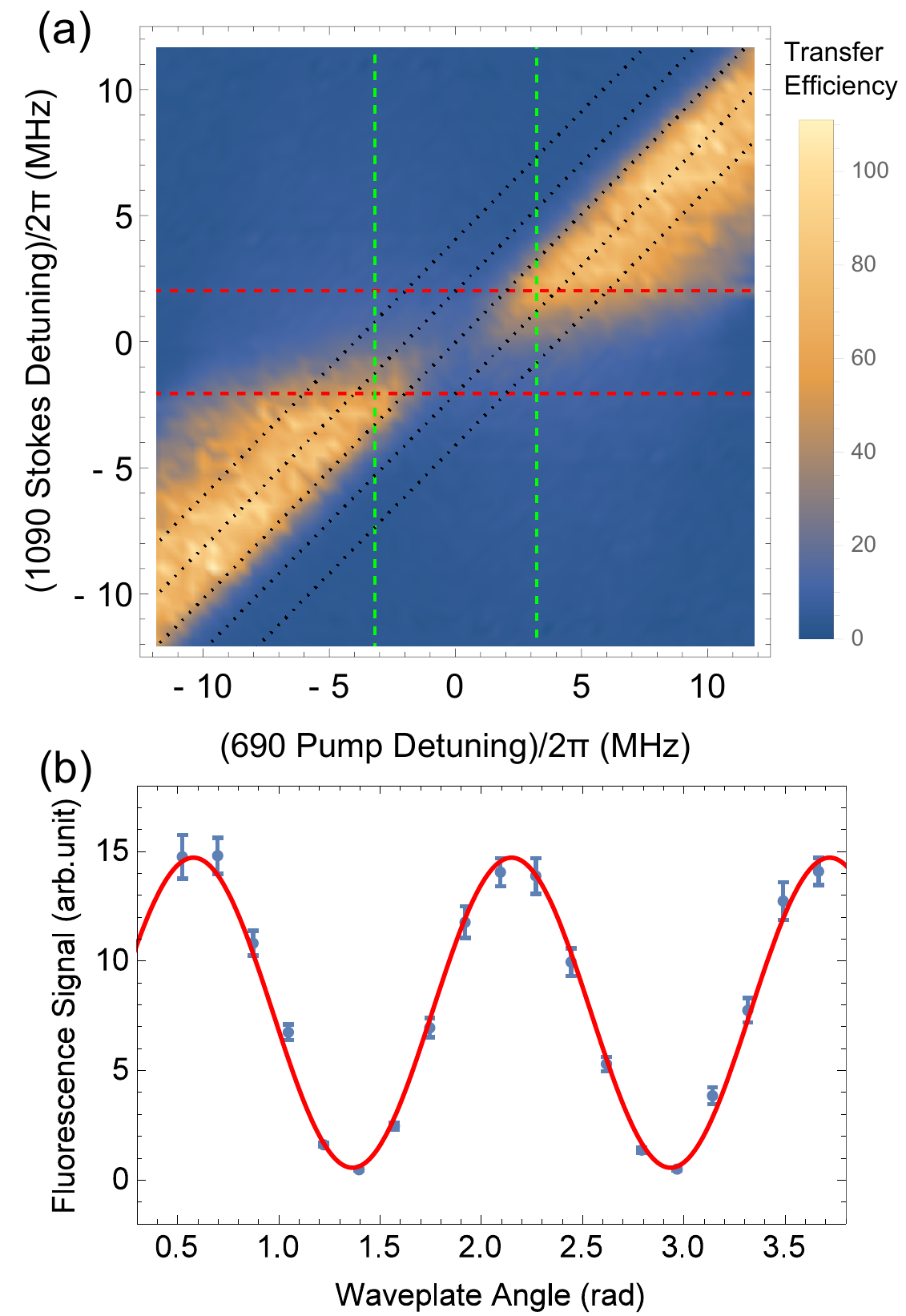}
\caption{(a) Density plot showing variation of the measured transfer efficiency with the detunings of the two lasers. Green (red) dashed lines are a guide to the eye, and indicate the extent of the pump (Stokes) one-photon resonance FWHM linewidth. Dotted black lines that indicate constant two-photon detunings are drawn at $\delta \in 2\pi \times \{-4,-2,0,+2,+4\}$ MHz.  (b) Spin analysis fringes showing coherent preparation of the aligned-spin electron state, data (blue) and sinusoidal fit (red line). \label{Fig4}}
\end{figure}

Figure \ref{Fig4}(a) shows the variation of the transfer efficiency with the detunings of the two lasers. This data is taken with both lasers at full power, with optimal overlap. As expected, STIRAP efficiency is very sensitive to two-photon detunings $\delta\neq 0$, but quite robust to one-photon detunings $\Delta \neq 0$. Unlike many other STIRAP systems, we obtain higher efficiency when running at one-photon detunings that are large compared to the Doppler linewidths of the pump and the Stokes beams. Transfer efficiency decreases substantially, to 10--20$\%$, when on one-photon resonance. We believe that this is due to sharp changes in the spatial intensity profiles of the laser beams across the large spatial extent (25 mm) of the molecular beam. On resonance, any excess laser intensity in the wings of the spatial intensity profiles (due e.g. to scatter or diffraction from apertures in the laser beam paths) can drive optical pumping, depleting the population of the initial state. This effect is accentuated by the relatively small transition dipole moment of the $|2\rangle \rightarrow |3\rangle$ Stokes transition, since population in the intermediary $|2\rangle$ state decays to other non-resonant molecular states (mostly to other rotational levels of the ground state $X$) with a much larger probability than to the desired state $|3\rangle$. Simulations of the molecule ensemble, when integrating the Schroedinger equation of the three-level system, show similar behavior when including Airy-like wings in the spatial intensity profiles of the lasers with amplitudes at the level of 10--20$\%$ of the maximum intensity, comparable to the ones observed in our laser beams. 

At one-photon detunings that are large compared to the one-photon Doppler linewidths of the pump and Stokes lasers, the two-photon profile  is asymmetric (Fig. \ref{Fig4}(a)): the transfer efficiency drops at a faster rate as the two-photon detuning  is tuned towards the pump one-photon resonance than when the two-photon detuning is tuned towards the Stokes one-photon resonance. This is general STIRAP behavior that occurs when at one-photon detunings that are large compared to $\Omega_{\rm eff}$ and when the Rabi frequencies are not equal \cite{Boradjiev2010}. In our case $\Omega_P/\Omega_S\approx 1.5$.

We verified that the STIRAP process directly populates a spin-aligned state $|3\rangle$ required for performing the spin precession measurement \cite{Kirilov2013}. Figure \ref{Fig4}(b) shows sinusoidal oscillations of the fluorescence signal characteristic of our spin analysis method, as the angle of the linear polarization of the readout beam is varied with a half-wave plate. The contrast of the spin analysis fringes is $93\pm2 \%$, comparable to that observed with the GEN I preparation method, $94 \pm 2 \%$ \cite{Baron2014}.

\section{Conclusion}

The reported results illustrate that a much higher fraction of the ThO molecules produced by our buffer gas beam source can be used for an electron EDM measurement than in ACME I \cite{Baron2014}.  STIRAP directly prepares the coherent superposition of ThO $H^3\Delta_1$ states that represents the initial spin-aligned state of our precession measurement and is approximately 12 times more efficient than the optical pumping method used in ACME I. Detecting fluorescence at a wavelength not used for the STIRAP excitation avoids stray light contamination while also allowing us to profit from the 2.5 times higher PMT quantum efficiency reported \footnote{Hamamatsu R7600U-300 and Hamamatsu R8900U-20 technical specifications.} for the new, shorter, 512 nm detection wavelength \cite{Kokkin2014}. In the course of this work, we also demonstrated that transferring fluorescence photons to the PMTs through light guides (rather than the fiber bundles used previously) increases the number of EDM-sensitive state molecules detected by a factor of 2.5.  These factors together illustrate that at least a 75 times higher fraction of the ThO molecules from our source can be used to perform a new electron EDM measurement using the metastable $H^3\Delta_1$ state of ThO. Other planned improvements in ACME (to be discussed elsewhere) could yield additional gains in the useful fraction of molecules and the source flux.

We have presented in detail the implementation of STIRAP in our system and contrasted it with the previously used optical pumping method of preparing the desired spin-aligned state. We have discussed the challenges of implementing STIRAP in ThO and described the methods we have used to overcome them. We have demonstrated that STIRAP can be applied with high transfer efficiency ($75\pm5 \%$) in systems characterized by low transition dipole moments, large volume of excitation, and with limited optical access. Finally, even higher transfer efficiencies could be achieved in the future with a higher Stokes laser power or improved laser spatial intensity profiles.

\begin{acknowledgments}
This work was performed as part of the ACME Collaboration, to whom we are grateful for its contributions, and was supported by the NSF.
\end{acknowledgments}

\bibliography{STIRAPPaper}

\begin{thebibliography}{43}%
\makeatletter
\providecommand \@ifxundefined [1]{%
 \@ifx{#1\undefined}
}%
\providecommand \@ifnum [1]{%
 \ifnum #1\expandafter \@firstoftwo
 \else \expandafter \@secondoftwo
 \fi
}%
\providecommand \@ifx [1]{%
 \ifx #1\expandafter \@firstoftwo
 \else \expandafter \@secondoftwo
 \fi
}%
\providecommand \natexlab [1]{#1}%
\providecommand \enquote  [1]{``#1''}%
\providecommand \bibnamefont  [1]{#1}%
\providecommand \bibfnamefont [1]{#1}%
\providecommand \citenamefont [1]{#1}%
\providecommand \href@noop [0]{\@secondoftwo}%
\providecommand \href [0]{\begingroup \@sanitize@url \@href}%
\providecommand \@href[1]{\@@startlink{#1}\@@href}%
\providecommand \@@href[1]{\endgroup#1\@@endlink}%
\providecommand \@sanitize@url [0]{\catcode `\\12\catcode `\$12\catcode
  `\&12\catcode `\#12\catcode `\^12\catcode `\_12\catcode `\%12\relax}%
\providecommand \@@startlink[1]{}%
\providecommand \@@endlink[0]{}%
\providecommand \url  [0]{\begingroup\@sanitize@url \@url }%
\providecommand \@url [1]{\endgroup\@href {#1}{\urlprefix }}%
\providecommand \urlprefix  [0]{URL }%
\providecommand \Eprint [0]{\href }%
\providecommand \doibase [0]{http://dx.doi.org/}%
\providecommand \selectlanguage [0]{\@gobble}%
\providecommand \bibinfo  [0]{\@secondoftwo}%
\providecommand \bibfield  [0]{\@secondoftwo}%
\providecommand \translation [1]{[#1]}%
\providecommand \BibitemOpen [0]{}%
\providecommand \bibitemStop [0]{}%
\providecommand \bibitemNoStop [0]{.\EOS\space}%
\providecommand \EOS [0]{\spacefactor3000\relax}%
\providecommand \BibitemShut  [1]{\csname bibitem#1\endcsname}%
\let\auto@bib@innerbib\@empty
\bibitem [{\citenamefont {Baron}\ \emph {et~al.}(2014)\citenamefont {Baron},
  \citenamefont {Campbell}, \citenamefont {DeMille}, \citenamefont {Doyle},
  \citenamefont {Gabrielse}, \citenamefont {Gurevich}, \citenamefont {Hess},
  \citenamefont {Hutzler}, \citenamefont {Kirilov}, \citenamefont {Kozyryev},
  \citenamefont {O'Leary}, \citenamefont {Panda}, \citenamefont {Parsons},
  \citenamefont {Petrik}, \citenamefont {Spaun}, \citenamefont {Vutha},\ and\
  \citenamefont {West}}]{Baron2014}%
  \BibitemOpen
  \bibfield  {author} {\bibinfo {author} {\bibfnamefont {J.}~\bibnamefont
  {Baron}}, \bibinfo {author} {\bibfnamefont {W.~C.}\ \bibnamefont {Campbell}},
  \bibinfo {author} {\bibfnamefont {D.}~\bibnamefont {DeMille}}, \bibinfo
  {author} {\bibfnamefont {J.~M.}\ \bibnamefont {Doyle}}, \bibinfo {author}
  {\bibfnamefont {G.}~\bibnamefont {Gabrielse}}, \bibinfo {author}
  {\bibfnamefont {Y.~V.}\ \bibnamefont {Gurevich}}, \bibinfo {author}
  {\bibfnamefont {P.~W.}\ \bibnamefont {Hess}}, \bibinfo {author}
  {\bibfnamefont {N.~R.}\ \bibnamefont {Hutzler}}, \bibinfo {author}
  {\bibfnamefont {E.}~\bibnamefont {Kirilov}}, \bibinfo {author} {\bibfnamefont
  {I.}~\bibnamefont {Kozyryev}}, \bibinfo {author} {\bibfnamefont {B.~R.}\
  \bibnamefont {O'Leary}}, \bibinfo {author} {\bibfnamefont {C.~D.}\
  \bibnamefont {Panda}}, \bibinfo {author} {\bibfnamefont {M.~F.}\ \bibnamefont
  {Parsons}}, \bibinfo {author} {\bibfnamefont {E.~S.}\ \bibnamefont {Petrik}},
  \bibinfo {author} {\bibfnamefont {B.}~\bibnamefont {Spaun}}, \bibinfo
  {author} {\bibfnamefont {A.~C.}\ \bibnamefont {Vutha}}, \ and\ \bibinfo
  {author} {\bibfnamefont {A.~D.}\ \bibnamefont {West}},\ }\href {\doibase
  10.1126/science.1248213} {\bibfield  {journal} {\bibinfo  {journal}
  {Science}\ }\textbf {\bibinfo {volume} {343}},\ \bibinfo {pages} {269}
  (\bibinfo {year} {2014})}\BibitemShut {NoStop}%
\bibitem [{Note1()}]{Note1}%
  \BibitemOpen
  \bibinfo {note} {Note that the limit reported here uses an updated value for
  $\protect \mathcal {E}_{\protect \rm eff}$ = 78 GV/cm, obtained from an
  unweighted mean of Refs. \cite {Skripnikov2015,Fleig2014}}\BibitemShut
  {NoStop}%
\bibitem [{\citenamefont {Hudson}\ \emph {et~al.}(2011)\citenamefont {Hudson},
  \citenamefont {Kara}, \citenamefont {Smallman}, \citenamefont {Sauer},
  \citenamefont {Tarbutt},\ and\ \citenamefont {Hinds}}]{Hudson2011}%
  \BibitemOpen
  \bibfield  {author} {\bibinfo {author} {\bibfnamefont {J.~J.}\ \bibnamefont
  {Hudson}}, \bibinfo {author} {\bibfnamefont {D.~M.}\ \bibnamefont {Kara}},
  \bibinfo {author} {\bibfnamefont {I.~J.}\ \bibnamefont {Smallman}}, \bibinfo
  {author} {\bibfnamefont {B.~E.}\ \bibnamefont {Sauer}}, \bibinfo {author}
  {\bibfnamefont {M.~R.}\ \bibnamefont {Tarbutt}}, \ and\ \bibinfo {author}
  {\bibfnamefont {E.~A.}\ \bibnamefont {Hinds}},\ }\href {\doibase
  10.1038/nature10104} {\bibfield  {journal} {\bibinfo  {journal} {Nature}\
  }\textbf {\bibinfo {volume} {473}},\ \bibinfo {pages} {493} (\bibinfo {year}
  {2011})}\BibitemShut {NoStop}%
\bibitem [{\citenamefont {Regan}\ \emph {et~al.}(2002)\citenamefont {Regan},
  \citenamefont {Commins}, \citenamefont {Schmidt},\ and\ \citenamefont
  {DeMille}}]{Regan2002}%
  \BibitemOpen
  \bibfield  {author} {\bibinfo {author} {\bibfnamefont {B.~C.}\ \bibnamefont
  {Regan}}, \bibinfo {author} {\bibfnamefont {E.~D.}\ \bibnamefont {Commins}},
  \bibinfo {author} {\bibfnamefont {C.~J.}\ \bibnamefont {Schmidt}}, \ and\
  \bibinfo {author} {\bibfnamefont {D.}~\bibnamefont {DeMille}},\ }\href
  {\doibase 10.1103/PhysRevLett.88.071805} {\bibfield  {journal} {\bibinfo
  {journal} {Phys. Rev. Lett.}\ }\textbf {\bibinfo {volume} {88}},\ \bibinfo
  {pages} {071805} (\bibinfo {year} {2002})}\BibitemShut {NoStop}%
\bibitem [{\citenamefont {Pospelov}\ and\ \citenamefont
  {Ritz}(2005)}]{Pospelov2005}%
  \BibitemOpen
  \bibfield  {author} {\bibinfo {author} {\bibfnamefont {M.}~\bibnamefont
  {Pospelov}}\ and\ \bibinfo {author} {\bibfnamefont {A.}~\bibnamefont
  {Ritz}},\ }\href {\doibase 10.1016/j.aop.2005.04.002} {\bibfield  {journal}
  {\bibinfo  {journal} {Ann. Phys.}\ }\textbf {\bibinfo {volume} {318}},\
  \bibinfo {pages} {119} (\bibinfo {year} {2005})}\BibitemShut {NoStop}%
\bibitem [{\citenamefont {Roberts}\ \emph {et~al.}(2009)\citenamefont
  {Roberts}, \citenamefont {Marciano},\ and\ \citenamefont
  {Eds.}}]{Roberts2009}%
  \BibitemOpen
  \bibfield  {author} {\bibinfo {author} {\bibfnamefont {B.~L.}\ \bibnamefont
  {Roberts}}, \bibinfo {author} {\bibfnamefont {W.~J.}\ \bibnamefont
  {Marciano}}, \ and\ \bibinfo {author} {\bibnamefont {Eds.}},\ }\href
  {http://arxiv.org/abs/hep-ex/0309010$\backslash$nhttp://www.worldscientific.com/doi/pdf/10.1142/9789814271844{\_}bmatter$\backslash$nhttp://scitation.aip.org/content/aip/proceeding/aipcp/10.1063/1.1664193}
  {\emph {\bibinfo {title} {World Scientific, Singapore}}}\ (\bibinfo {year}
  {2009})\BibitemShut {NoStop}%
\bibitem [{\citenamefont {Kirilov}\ \emph {et~al.}(2013)\citenamefont
  {Kirilov}, \citenamefont {Campbell}, \citenamefont {Doyle}, \citenamefont
  {Gabrielse}, \citenamefont {Gurevich}, \citenamefont {Hess}, \citenamefont
  {Hutzler}, \citenamefont {O'Leary}, \citenamefont {Petrik}, \citenamefont
  {Spaun}, \citenamefont {Vutha},\ and\ \citenamefont {DeMille}}]{Kirilov2013}%
  \BibitemOpen
  \bibfield  {author} {\bibinfo {author} {\bibfnamefont {E.}~\bibnamefont
  {Kirilov}}, \bibinfo {author} {\bibfnamefont {W.~C.}\ \bibnamefont
  {Campbell}}, \bibinfo {author} {\bibfnamefont {J.~M.}\ \bibnamefont {Doyle}},
  \bibinfo {author} {\bibfnamefont {G.}~\bibnamefont {Gabrielse}}, \bibinfo
  {author} {\bibfnamefont {Y.~V.}\ \bibnamefont {Gurevich}}, \bibinfo {author}
  {\bibfnamefont {P.~W.}\ \bibnamefont {Hess}}, \bibinfo {author}
  {\bibfnamefont {N.~R.}\ \bibnamefont {Hutzler}}, \bibinfo {author}
  {\bibfnamefont {B.~R.}\ \bibnamefont {O'Leary}}, \bibinfo {author}
  {\bibfnamefont {E.}~\bibnamefont {Petrik}}, \bibinfo {author} {\bibfnamefont
  {B.}~\bibnamefont {Spaun}}, \bibinfo {author} {\bibfnamefont {A.~C.}\
  \bibnamefont {Vutha}}, \ and\ \bibinfo {author} {\bibfnamefont
  {D.}~\bibnamefont {DeMille}},\ }\href {\doibase 10.1103/PhysRevA.88.013844}
  {\bibfield  {journal} {\bibinfo  {journal} {Phys. Rev. A}\ }\textbf {\bibinfo
  {volume} {88}},\ \bibinfo {pages} {1} (\bibinfo {year} {2013})}\BibitemShut
  {NoStop}%
\bibitem [{\citenamefont {Hutzler}\ \emph {et~al.}(2011)\citenamefont
  {Hutzler}, \citenamefont {Parsons}, \citenamefont {Gurevich}, \citenamefont
  {Hess}, \citenamefont {Petrik}, \citenamefont {Spaun}, \citenamefont {Vutha},
  \citenamefont {DeMille}, \citenamefont {Gabrielse},\ and\ \citenamefont
  {Doyle}}]{Hutzler2011}%
  \BibitemOpen
  \bibfield  {author} {\bibinfo {author} {\bibfnamefont {N.~R.}\ \bibnamefont
  {Hutzler}}, \bibinfo {author} {\bibfnamefont {M.}~\bibnamefont {Parsons}},
  \bibinfo {author} {\bibfnamefont {Y.~V.}\ \bibnamefont {Gurevich}}, \bibinfo
  {author} {\bibfnamefont {P.~W.}\ \bibnamefont {Hess}}, \bibinfo {author}
  {\bibfnamefont {E.}~\bibnamefont {Petrik}}, \bibinfo {author} {\bibfnamefont
  {B.}~\bibnamefont {Spaun}}, \bibinfo {author} {\bibfnamefont {A.~C.}\
  \bibnamefont {Vutha}}, \bibinfo {author} {\bibfnamefont {D.}~\bibnamefont
  {DeMille}}, \bibinfo {author} {\bibfnamefont {G.}~\bibnamefont {Gabrielse}},
  \ and\ \bibinfo {author} {\bibfnamefont {J.~M.}\ \bibnamefont {Doyle}},\
  }\href {\doibase 10.1039/c1cp20901a} {\bibfield  {journal} {\bibinfo
  {journal} {Phys. Chem. Chem. Phys.}\ }\textbf {\bibinfo {volume} {13}},\
  \bibinfo {pages} {18976} (\bibinfo {year} {2011})}\BibitemShut {NoStop}%
\bibitem [{\citenamefont {Skripnikov}\ and\ \citenamefont
  {Titov}(2015)}]{Skripnikov2015}%
  \BibitemOpen
  \bibfield  {author} {\bibinfo {author} {\bibfnamefont {L.~V.}\ \bibnamefont
  {Skripnikov}}\ and\ \bibinfo {author} {\bibfnamefont {A.~V.}\ \bibnamefont
  {Titov}},\ }\href {\doibase 10.1063/1.4904877} {\bibfield  {journal}
  {\bibinfo  {journal} {J. Chem. Phys.}\ }\textbf {\bibinfo {volume} {142}},\
  \bibinfo {pages} {024301} (\bibinfo {year} {2015})}\BibitemShut {NoStop}%
\bibitem [{\citenamefont {Fleig}\ and\ \citenamefont
  {Nayak}(2014)}]{Fleig2014}%
  \BibitemOpen
  \bibfield  {author} {\bibinfo {author} {\bibfnamefont {T.}~\bibnamefont
  {Fleig}}\ and\ \bibinfo {author} {\bibfnamefont {M.~K.}\ \bibnamefont
  {Nayak}},\ }\href {\doibase 10.1016/j.jms.2014.03.017} {\bibfield  {journal}
  {\bibinfo  {journal} {J. Mol. Spectrosc.}\ }\textbf {\bibinfo {volume}
  {300}},\ \bibinfo {pages} {16} (\bibinfo {year} {2014})}\BibitemShut
  {NoStop}%
\bibitem [{\citenamefont {Meyer}\ and\ \citenamefont {Bohn}(2008)}]{Meyer2008}%
  \BibitemOpen
  \bibfield  {author} {\bibinfo {author} {\bibfnamefont {E.~R.}\ \bibnamefont
  {Meyer}}\ and\ \bibinfo {author} {\bibfnamefont {J.~L.}\ \bibnamefont
  {Bohn}},\ }\href {\doibase 10.1103/PhysRevA.78.010502} {\bibfield  {journal}
  {\bibinfo  {journal} {Phys. Rev. A}\ }\textbf {\bibinfo {volume} {78}},\
  \bibinfo {pages} {1} (\bibinfo {year} {2008})}\BibitemShut {NoStop}%
\bibitem [{\citenamefont {Bickman}\ \emph {et~al.}(2009)\citenamefont
  {Bickman}, \citenamefont {Hamilton}, \citenamefont {Jiang},\ and\
  \citenamefont {DeMille}}]{Bickman2009}%
  \BibitemOpen
  \bibfield  {author} {\bibinfo {author} {\bibfnamefont {S.}~\bibnamefont
  {Bickman}}, \bibinfo {author} {\bibfnamefont {P.}~\bibnamefont {Hamilton}},
  \bibinfo {author} {\bibfnamefont {Y.}~\bibnamefont {Jiang}}, \ and\ \bibinfo
  {author} {\bibfnamefont {D.}~\bibnamefont {DeMille}},\ }\href {\doibase
  10.1103/PhysRevA.80.023418} {\bibfield  {journal} {\bibinfo  {journal} {Phys.
  Rev. A}\ }\textbf {\bibinfo {volume} {80}},\ \bibinfo {pages} {1} (\bibinfo
  {year} {2009})}\BibitemShut {NoStop}%
\bibitem [{\citenamefont {Petrov}\ \emph {et~al.}(2014)\citenamefont {Petrov},
  \citenamefont {Skripnikov}, \citenamefont {Titov}, \citenamefont {Hutzler},
  \citenamefont {Hess}, \citenamefont {O'Leary}, \citenamefont {Spaun},
  \citenamefont {DeMille}, \citenamefont {Gabrielse},\ and\ \citenamefont
  {Doyle}}]{Petrov2014}%
  \BibitemOpen
  \bibfield  {author} {\bibinfo {author} {\bibfnamefont {A.~N.}\ \bibnamefont
  {Petrov}}, \bibinfo {author} {\bibfnamefont {L.~V.}\ \bibnamefont
  {Skripnikov}}, \bibinfo {author} {\bibfnamefont {A.~V.}\ \bibnamefont
  {Titov}}, \bibinfo {author} {\bibfnamefont {N.~R.}\ \bibnamefont {Hutzler}},
  \bibinfo {author} {\bibfnamefont {P.~W.}\ \bibnamefont {Hess}}, \bibinfo
  {author} {\bibfnamefont {B.~R.}\ \bibnamefont {O'Leary}}, \bibinfo {author}
  {\bibfnamefont {B.}~\bibnamefont {Spaun}}, \bibinfo {author} {\bibfnamefont
  {D.}~\bibnamefont {DeMille}}, \bibinfo {author} {\bibfnamefont
  {G.}~\bibnamefont {Gabrielse}}, \ and\ \bibinfo {author} {\bibfnamefont
  {J.~M.}\ \bibnamefont {Doyle}},\ }\href {\doibase 10.1103/PhysRevA.89.062505}
  {\bibfield  {journal} {\bibinfo  {journal} {Phys. Rev. A}\ }\textbf {\bibinfo
  {volume} {89}},\ \bibinfo {pages} {062505} (\bibinfo {year}
  {2014})}\BibitemShut {NoStop}%
\bibitem [{\citenamefont {Vutha}\ \emph {et~al.}(2010)\citenamefont {Vutha},
  \citenamefont {Campbell}, \citenamefont {Gurevich}, \citenamefont {Hutzler},
  \citenamefont {Parsons}, \citenamefont {Patterson}, \citenamefont {Petrik},
  \citenamefont {Spaun}, \citenamefont {Doyle}, \citenamefont {Gabrielse},\
  and\ \citenamefont {DeMille}}]{Vutha2010}%
  \BibitemOpen
  \bibfield  {author} {\bibinfo {author} {\bibfnamefont {A.~C.}\ \bibnamefont
  {Vutha}}, \bibinfo {author} {\bibfnamefont {W.~C.}\ \bibnamefont {Campbell}},
  \bibinfo {author} {\bibfnamefont {Y.~V.}\ \bibnamefont {Gurevich}}, \bibinfo
  {author} {\bibfnamefont {N.~R.}\ \bibnamefont {Hutzler}}, \bibinfo {author}
  {\bibfnamefont {M.}~\bibnamefont {Parsons}}, \bibinfo {author} {\bibfnamefont
  {D.}~\bibnamefont {Patterson}}, \bibinfo {author} {\bibfnamefont
  {E.}~\bibnamefont {Petrik}}, \bibinfo {author} {\bibfnamefont
  {B.}~\bibnamefont {Spaun}}, \bibinfo {author} {\bibfnamefont {J.~M.}\
  \bibnamefont {Doyle}}, \bibinfo {author} {\bibfnamefont {G.}~\bibnamefont
  {Gabrielse}}, \ and\ \bibinfo {author} {\bibfnamefont {D.}~\bibnamefont
  {DeMille}},\ }\href {\doibase 10.1088/0953-4075/43/7/074007} {\bibfield
  {journal} {\bibinfo  {journal} {J. Phys. B}\ }\textbf {\bibinfo {volume}
  {43}},\ \bibinfo {pages} {074007} (\bibinfo {year} {2010})},\ \Eprint
  {http://arxiv.org/abs/0908.2412} {arXiv:0908.2412} \BibitemShut {NoStop}%
\bibitem [{\citenamefont {Bergmann}\ \emph {et~al.}(1998)\citenamefont
  {Bergmann}, \citenamefont {Theuer},\ and\ \citenamefont
  {Shore}}]{Bergmann1998}%
  \BibitemOpen
  \bibfield  {author} {\bibinfo {author} {\bibfnamefont {K.}~\bibnamefont
  {Bergmann}}, \bibinfo {author} {\bibfnamefont {H.}~\bibnamefont {Theuer}}, \
  and\ \bibinfo {author} {\bibfnamefont {B.~W.}\ \bibnamefont {Shore}},\ }\href
  {\doibase 10.1103/RevModPhys.70.1003} {\bibfield  {journal} {\bibinfo
  {journal} {Rev. Mod. Phys.}\ }\textbf {\bibinfo {volume} {70}},\ \bibinfo
  {pages} {1003} (\bibinfo {year} {1998})}\BibitemShut {NoStop}%
\bibitem [{\citenamefont {Gaubatz}\ \emph {et~al.}(1990)\citenamefont
  {Gaubatz}, \citenamefont {Rudecki}, \citenamefont {Schiemann},\ and\
  \citenamefont {Bergmann}}]{Gaubatz1990}%
  \BibitemOpen
  \bibfield  {author} {\bibinfo {author} {\bibfnamefont {U.}~\bibnamefont
  {Gaubatz}}, \bibinfo {author} {\bibfnamefont {P.}~\bibnamefont {Rudecki}},
  \bibinfo {author} {\bibfnamefont {S.}~\bibnamefont {Schiemann}}, \ and\
  \bibinfo {author} {\bibfnamefont {K.}~\bibnamefont {Bergmann}},\ }\href
  {\doibase 10.1063/1.458514} {\bibfield  {journal} {\bibinfo  {journal} {J
  .Chem. Phys.}\ }\textbf {\bibinfo {volume} {92}},\ \bibinfo {pages} {5363}
  (\bibinfo {year} {1990})}\BibitemShut {NoStop}%
\bibitem [{\citenamefont {Ni}\ \emph {et~al.}(2008)\citenamefont {Ni},
  \citenamefont {Ospelkaus}, \citenamefont {de~Miranda}, \citenamefont {Pe'er},
  \citenamefont {Neyenhuis}, \citenamefont {Zirbel}, \citenamefont
  {Kotochigova}, \citenamefont {Julienne}, \citenamefont {Jin},\ and\
  \citenamefont {Ye}}]{Ni2008}%
  \BibitemOpen
  \bibfield  {author} {\bibinfo {author} {\bibfnamefont {K.-K.}\ \bibnamefont
  {Ni}}, \bibinfo {author} {\bibfnamefont {S.}~\bibnamefont {Ospelkaus}},
  \bibinfo {author} {\bibfnamefont {M.~H.~G.}\ \bibnamefont {de~Miranda}},
  \bibinfo {author} {\bibfnamefont {A.}~\bibnamefont {Pe'er}}, \bibinfo
  {author} {\bibfnamefont {B.}~\bibnamefont {Neyenhuis}}, \bibinfo {author}
  {\bibfnamefont {J.~J.}\ \bibnamefont {Zirbel}}, \bibinfo {author}
  {\bibfnamefont {S.}~\bibnamefont {Kotochigova}}, \bibinfo {author}
  {\bibfnamefont {P.~S.}\ \bibnamefont {Julienne}}, \bibinfo {author}
  {\bibfnamefont {D.~S.}\ \bibnamefont {Jin}}, \ and\ \bibinfo {author}
  {\bibfnamefont {J.}~\bibnamefont {Ye}},\ }\href {\doibase
  10.1126/science.1163861} {\bibfield  {journal} {\bibinfo  {journal}
  {Science}\ }\textbf {\bibinfo {volume} {322}},\ \bibinfo {pages} {231}
  (\bibinfo {year} {2008})}\BibitemShut {NoStop}%
\bibitem [{\citenamefont {Danzl}\ \emph {et~al.}(2008)\citenamefont {Danzl},
  \citenamefont {Haller}, \citenamefont {Gustavsson}, \citenamefont {Mark},
  \citenamefont {Hart}, \citenamefont {Bouloufa}, \citenamefont {Dulieu},
  \citenamefont {Ritsch},\ and\ \citenamefont {Nagerl}}]{Danzl2008}%
  \BibitemOpen
  \bibfield  {author} {\bibinfo {author} {\bibfnamefont {J.~G.}\ \bibnamefont
  {Danzl}}, \bibinfo {author} {\bibfnamefont {E.}~\bibnamefont {Haller}},
  \bibinfo {author} {\bibfnamefont {M.}~\bibnamefont {Gustavsson}}, \bibinfo
  {author} {\bibfnamefont {M.~J.}\ \bibnamefont {Mark}}, \bibinfo {author}
  {\bibfnamefont {R.}~\bibnamefont {Hart}}, \bibinfo {author} {\bibfnamefont
  {N.}~\bibnamefont {Bouloufa}}, \bibinfo {author} {\bibfnamefont
  {O.}~\bibnamefont {Dulieu}}, \bibinfo {author} {\bibfnamefont
  {H.}~\bibnamefont {Ritsch}}, \ and\ \bibinfo {author} {\bibfnamefont {H.~C.}\
  \bibnamefont {Nagerl}},\ }\href {\doibase DOI 10.1126/science.1159909}
  {\bibfield  {journal} {\bibinfo  {journal} {Science}\ }\textbf {\bibinfo
  {volume} {321}},\ \bibinfo {pages} {1062} (\bibinfo {year}
  {2008})}\BibitemShut {NoStop}%
\bibitem [{\citenamefont {Parkins}\ \emph {et~al.}(1993)\citenamefont
  {Parkins}, \citenamefont {Marte}, \citenamefont {Zoller},\ and\ \citenamefont
  {Kimble}}]{Parkins1993}%
  \BibitemOpen
  \bibfield  {author} {\bibinfo {author} {\bibfnamefont {A.~S.}\ \bibnamefont
  {Parkins}}, \bibinfo {author} {\bibfnamefont {P.}~\bibnamefont {Marte}},
  \bibinfo {author} {\bibfnamefont {P.}~\bibnamefont {Zoller}}, \ and\ \bibinfo
  {author} {\bibfnamefont {H.~J.}\ \bibnamefont {Kimble}},\ }\href {\doibase
  10.1103/PhysRevLett.71.3095} {\bibfield  {journal} {\bibinfo  {journal}
  {Phys. Rev. Lett.}\ }\textbf {\bibinfo {volume} {71}},\ \bibinfo {pages}
  {3095} (\bibinfo {year} {1993})}\BibitemShut {NoStop}%
\bibitem [{\citenamefont {Hennrich}\ \emph {et~al.}(2000)\citenamefont
  {Hennrich}, \citenamefont {Legero}, \citenamefont {Kuhn},\ and\ \citenamefont
  {Rempe}}]{Hennrich2000}%
  \BibitemOpen
  \bibfield  {author} {\bibinfo {author} {\bibfnamefont {M.}~\bibnamefont
  {Hennrich}}, \bibinfo {author} {\bibfnamefont {T.}~\bibnamefont {Legero}},
  \bibinfo {author} {\bibfnamefont {A.}~\bibnamefont {Kuhn}}, \ and\ \bibinfo
  {author} {\bibfnamefont {G.}~\bibnamefont {Rempe}},\ }\href {\doibase
  10.1103/PhysRevLett.85.4872} {\bibfield  {journal} {\bibinfo  {journal}
  {Phys. Rev. Lett.}\ }\textbf {\bibinfo {volume} {85}},\ \bibinfo {pages}
  {4872} (\bibinfo {year} {2000})}\BibitemShut {NoStop}%
\bibitem [{\citenamefont {S{\o}rensen}\ \emph {et~al.}(2006)\citenamefont
  {S{\o}rensen}, \citenamefont {M{\o}ller}, \citenamefont {Iversen},
  \citenamefont {Thomsen}, \citenamefont {Jensen}, \citenamefont {Staanum},
  \citenamefont {Voigt},\ and\ \citenamefont {Drewsen}}]{Sorensen2006}%
  \BibitemOpen
  \bibfield  {author} {\bibinfo {author} {\bibfnamefont {J.}~\bibnamefont
  {S{\o}rensen}}, \bibinfo {author} {\bibfnamefont {D.}~\bibnamefont
  {M{\o}ller}}, \bibinfo {author} {\bibfnamefont {T.}~\bibnamefont {Iversen}},
  \bibinfo {author} {\bibfnamefont {J.~B.}\ \bibnamefont {Thomsen}}, \bibinfo
  {author} {\bibfnamefont {F.}~\bibnamefont {Jensen}}, \bibinfo {author}
  {\bibfnamefont {P.}~\bibnamefont {Staanum}}, \bibinfo {author} {\bibfnamefont
  {D.}~\bibnamefont {Voigt}}, \ and\ \bibinfo {author} {\bibfnamefont
  {M.}~\bibnamefont {Drewsen}},\ }\href {\doibase 10.1088/1367-2630/8/11/261}
  {\bibfield  {journal} {\bibinfo  {journal} {New J. Phys.}\ }\textbf {\bibinfo
  {volume} {8}},\ \bibinfo {pages} {261} (\bibinfo {year} {2006})}\BibitemShut
  {NoStop}%
\bibitem [{\citenamefont {M{\o}ller}\ \emph {et~al.}(2007)\citenamefont
  {M{\o}ller}, \citenamefont {S{\o}rensen}, \citenamefont {Thomsen},\ and\
  \citenamefont {Drewsen}}]{Moller2007}%
  \BibitemOpen
  \bibfield  {author} {\bibinfo {author} {\bibfnamefont {D.}~\bibnamefont
  {M{\o}ller}}, \bibinfo {author} {\bibfnamefont {J.}~\bibnamefont
  {S{\o}rensen}}, \bibinfo {author} {\bibfnamefont {J.}~\bibnamefont
  {Thomsen}}, \ and\ \bibinfo {author} {\bibfnamefont {M.}~\bibnamefont
  {Drewsen}},\ }\href {\doibase 10.1103/PhysRevA.76.062321} {\bibfield
  {journal} {\bibinfo  {journal} {Phys. Rev. A}\ }\textbf {\bibinfo {volume}
  {76}},\ \bibinfo {pages} {062321} (\bibinfo {year} {2007})}\BibitemShut
  {NoStop}%
\bibitem [{\citenamefont {Yatsenko}\ \emph {et~al.}(2014)\citenamefont
  {Yatsenko}, \citenamefont {Shore},\ and\ \citenamefont
  {Bergmann}}]{Yatsenko2014}%
  \BibitemOpen
  \bibfield  {author} {\bibinfo {author} {\bibfnamefont {L.~P.}\ \bibnamefont
  {Yatsenko}}, \bibinfo {author} {\bibfnamefont {B.~W.}\ \bibnamefont {Shore}},
  \ and\ \bibinfo {author} {\bibfnamefont {K.}~\bibnamefont {Bergmann}},\
  }\href {\doibase 10.1103/PhysRevA.89.013831} {\bibfield  {journal} {\bibinfo
  {journal} {Phys. Rev. A}\ }\textbf {\bibinfo {volume} {89}},\ \bibinfo
  {pages} {1} (\bibinfo {year} {2014})}\BibitemShut {NoStop}%
\bibitem [{\citenamefont {Kuklinski}\ \emph {et~al.}(1989)\citenamefont
  {Kuklinski}, \citenamefont {Gaubatz}, \citenamefont {Hioe},\ and\
  \citenamefont {Bergmann}}]{Kuklinski1989}%
  \BibitemOpen
  \bibfield  {author} {\bibinfo {author} {\bibfnamefont {J.~R.}\ \bibnamefont
  {Kuklinski}}, \bibinfo {author} {\bibfnamefont {U.}~\bibnamefont {Gaubatz}},
  \bibinfo {author} {\bibfnamefont {F.~T.}\ \bibnamefont {Hioe}}, \ and\
  \bibinfo {author} {\bibfnamefont {K.}~\bibnamefont {Bergmann}},\ }\href
  {\doibase 10.1103/PhysRevA.40.6741} {\bibfield  {journal} {\bibinfo
  {journal} {Phys. Rev. A}\ }\textbf {\bibinfo {volume} {40}},\ \bibinfo
  {pages} {6741} (\bibinfo {year} {1989})}\BibitemShut {NoStop}%
\bibitem [{Note2()}]{Note2}%
  \BibitemOpen
  \bibinfo {note} {This efficiency number uses an updated calculation, which
  includes all paths of decay from the excited $A$ state, using data from the
  measurement discussed in Ref. \cite {Spaun2014} \label {note1}}\BibitemShut
  {NoStop}%
\bibitem [{\citenamefont {Hess}(2014)}]{Hess2014}%
  \BibitemOpen
  \bibfield  {author} {\bibinfo {author} {\bibfnamefont {P.~W.}\ \bibnamefont
  {Hess}},\ }\href@noop {} {\bibfield  {journal} {\bibinfo  {journal} {PhD
  thesis, Harvard University}\ } (\bibinfo {year} {2014})}\BibitemShut
  {NoStop}%
\bibitem [{\citenamefont {Kokkin}\ \emph {et~al.}(2014)\citenamefont {Kokkin},
  \citenamefont {Steimle},\ and\ \citenamefont {DeMille}}]{Kokkin2014}%
  \BibitemOpen
  \bibfield  {author} {\bibinfo {author} {\bibfnamefont {D.~L.}\ \bibnamefont
  {Kokkin}}, \bibinfo {author} {\bibfnamefont {T.~C.}\ \bibnamefont {Steimle}},
  \ and\ \bibinfo {author} {\bibfnamefont {D.}~\bibnamefont {DeMille}},\ }\href
  {\doibase 10.1103/PhysRevA.90.062503} {\bibfield  {journal} {\bibinfo
  {journal} {Phys. Rev. A}\ }\textbf {\bibinfo {volume} {90}},\ \bibinfo
  {pages} {1} (\bibinfo {year} {2014})}\BibitemShut {NoStop}%
\bibitem [{\citenamefont {Kokkin}\ \emph {et~al.}(2015)\citenamefont {Kokkin},
  \citenamefont {Steimle},\ and\ \citenamefont {DeMille}}]{Kokkin2015}%
  \BibitemOpen
  \bibfield  {author} {\bibinfo {author} {\bibfnamefont {D.~L.}\ \bibnamefont
  {Kokkin}}, \bibinfo {author} {\bibfnamefont {T.~C.}\ \bibnamefont {Steimle}},
  \ and\ \bibinfo {author} {\bibfnamefont {D.}~\bibnamefont {DeMille}},\ }\href
  {\doibase 10.1103/PhysRevA.91.042508} {\bibfield  {journal} {\bibinfo
  {journal} {Phys. Rev. A}\ }\textbf {\bibinfo {volume} {91}},\ \bibinfo
  {pages} {42508} (\bibinfo {year} {2015})}\BibitemShut {NoStop}%
\bibitem [{\citenamefont {Edvinsson}\ \emph {et~al.}(1968)\citenamefont
  {Edvinsson}, \citenamefont {Bornstedt},\ and\ \citenamefont
  {Nylen}}]{Edvinsson1968}%
  \BibitemOpen
  \bibfield  {author} {\bibinfo {author} {\bibfnamefont {G.}~\bibnamefont
  {Edvinsson}}, \bibinfo {author} {\bibfnamefont {A.}~\bibnamefont
  {Bornstedt}}, \ and\ \bibinfo {author} {\bibfnamefont {P.}~\bibnamefont
  {Nylen}},\ }\href@noop {} {\bibfield  {journal} {\bibinfo  {journal} {Ark.
  Fys.}\ }\textbf {\bibinfo {volume} {38}},\ \bibinfo {pages} {193} (\bibinfo
  {year} {1968})}\BibitemShut {NoStop}%
\bibitem [{Note3()}]{Note3}%
  \BibitemOpen
  \bibinfo {note} {All four lasers are Toptica DL pro with antireflection
  (AR)-coated diodes}\BibitemShut {NoStop}%
\bibitem [{\citenamefont {Drever}\ \emph {et~al.}(1983)\citenamefont {Drever},
  \citenamefont {Hall}, \citenamefont {Kowalski}, \citenamefont {Hough},
  \citenamefont {Ford}, \citenamefont {Munley},\ and\ \citenamefont
  {Ward}}]{Drever1983}%
  \BibitemOpen
  \bibfield  {author} {\bibinfo {author} {\bibfnamefont {R.~W.~P.}\
  \bibnamefont {Drever}}, \bibinfo {author} {\bibfnamefont {J.~L.}\
  \bibnamefont {Hall}}, \bibinfo {author} {\bibfnamefont {F.~V.}\ \bibnamefont
  {Kowalski}}, \bibinfo {author} {\bibfnamefont {J.}~\bibnamefont {Hough}},
  \bibinfo {author} {\bibfnamefont {G.~M.}\ \bibnamefont {Ford}}, \bibinfo
  {author} {\bibfnamefont {A.~J.}\ \bibnamefont {Munley}}, \ and\ \bibinfo
  {author} {\bibfnamefont {H.}~\bibnamefont {Ward}},\ }\href {\doibase
  10.1007/BF00702605} {\bibfield  {journal} {\bibinfo  {journal} {Appl. Phys.
  B}\ }\textbf {\bibinfo {volume} {31}},\ \bibinfo {pages} {97} (\bibinfo
  {year} {1983})}\BibitemShut {NoStop}%
\bibitem [{Note4()}]{Note4}%
  \BibitemOpen
  \bibinfo {note} {Commercial MSquared Solstis 4000 system.}\BibitemShut
  {Stop}%
\bibitem [{\citenamefont {Schunemann}\ \emph {et~al.}(1999)\citenamefont
  {Schunemann}, \citenamefont {Engler}, \citenamefont {Grimm}, \citenamefont
  {Weidemuller},\ and\ \citenamefont {Zielonkowski}}]{Schunemann1999}%
  \BibitemOpen
  \bibfield  {author} {\bibinfo {author} {\bibfnamefont {U.}~\bibnamefont
  {Schunemann}}, \bibinfo {author} {\bibfnamefont {H.}~\bibnamefont {Engler}},
  \bibinfo {author} {\bibfnamefont {R.}~\bibnamefont {Grimm}}, \bibinfo
  {author} {\bibfnamefont {M.}~\bibnamefont {Weidemuller}}, \ and\ \bibinfo
  {author} {\bibfnamefont {M.}~\bibnamefont {Zielonkowski}},\ }\href {\doibase
  10.1063/1.1149573} {\bibfield  {journal} {\bibinfo  {journal} {Rev. Sci.
  Instrum.}\ }\textbf {\bibinfo {volume} {70}},\ \bibinfo {pages} {242}
  (\bibinfo {year} {1999})}\BibitemShut {NoStop}%
\bibitem [{\citenamefont {Romanenko}\ and\ \citenamefont
  {Yatsenko}(1997)}]{Romanenko1997}%
  \BibitemOpen
  \bibfield  {author} {\bibinfo {author} {\bibfnamefont {V.}~\bibnamefont
  {Romanenko}}\ and\ \bibinfo {author} {\bibfnamefont {L.}~\bibnamefont
  {Yatsenko}},\ }\href {\doibase 10.1016/S0030-4018(97)00152-1} {\bibfield
  {journal} {\bibinfo  {journal} {Opt. Commun.}\ }\textbf {\bibinfo {volume}
  {140}},\ \bibinfo {pages} {231} (\bibinfo {year} {1997})}\BibitemShut
  {NoStop}%
\bibitem [{Note5()}]{Note5}%
  \BibitemOpen
  \bibinfo {note} {The power spectral density of the phase follows a Lorentzian
  distribution. In the limit that the integral of the phase power spectral
  density is small compared to 1, the laser field spectrum consists
  approximately of a delta function at the center frequency that contains most
  of the power and a Lorentzian pedestal elsewhere.}\BibitemShut {Stop}%
\bibitem [{\citenamefont {Halfmann}\ \emph {et~al.}(2003)\citenamefont
  {Halfmann}, \citenamefont {Rickes}, \citenamefont {Vitanov},\ and\
  \citenamefont {Bergmann}}]{Halfmann2003}%
  \BibitemOpen
  \bibfield  {author} {\bibinfo {author} {\bibfnamefont {T.}~\bibnamefont
  {Halfmann}}, \bibinfo {author} {\bibfnamefont {T.}~\bibnamefont {Rickes}},
  \bibinfo {author} {\bibfnamefont {N.~V.}\ \bibnamefont {Vitanov}}, \ and\
  \bibinfo {author} {\bibfnamefont {K.}~\bibnamefont {Bergmann}},\ }\href
  {\doibase 10.1016/S0030-4018(03)01368-3} {\bibfield  {journal} {\bibinfo
  {journal} {Opt. Commun.}\ }\textbf {\bibinfo {volume} {220}},\ \bibinfo
  {pages} {353} (\bibinfo {year} {2003})}\BibitemShut {NoStop}%
\bibitem [{\citenamefont {Spaun}(2014)}]{Spaun2014}%
  \BibitemOpen
  \bibfield  {author} {\bibinfo {author} {\bibfnamefont {B.~N.}\ \bibnamefont
  {Spaun}},\ }\href@noop {} {\bibfield  {journal} {\bibinfo  {journal} {PhD
  thesis, Harvard University}\ } (\bibinfo {year} {2014})}\BibitemShut
  {NoStop}%
\bibitem [{Note6()}]{Note6}%
  \BibitemOpen
  \bibinfo {note} {A Neon buffer gas flow of 40 SCCM was used in these
  measurements.}\BibitemShut {Stop}%
\bibitem [{Note7()}]{Note7}%
  \BibitemOpen
  \bibinfo {note} {A vibrational branching from ($H,v=0$) to ($X,v=0$) of $\sim
  94\%$ is calculated using the method of Ref. \cite {Nicholls1981} and
  spectroscopic data from Ref. \cite {Edvinsson1985}. Rotational branching from
  $(H,J=1,M=\pm 1 )$ to $(X,J=0)$ is $1/3$.}\BibitemShut {Stop}%
\bibitem [{\citenamefont {Boradjiev}\ and\ \citenamefont
  {Vitanov}(2010)}]{Boradjiev2010}%
  \BibitemOpen
  \bibfield  {author} {\bibinfo {author} {\bibfnamefont {I.~I.}\ \bibnamefont
  {Boradjiev}}\ and\ \bibinfo {author} {\bibfnamefont {N.~V.}\ \bibnamefont
  {Vitanov}},\ }\href {\doibase 10.1103/PhysRevA.81.053415} {\bibfield
  {journal} {\bibinfo  {journal} {Phys. Rev. A}\ }\textbf {\bibinfo {volume}
  {81}},\ \bibinfo {pages} {1} (\bibinfo {year} {2010})}\BibitemShut {NoStop}%
\bibitem [{Note8()}]{Note8}%
  \BibitemOpen
  \bibinfo {note} {Hamamatsu R7600U-300 and Hamamatsu R8900U-20 technical
  specifications.}\BibitemShut {Stop}%
\bibitem [{\citenamefont {Nicholls}(1981)}]{Nicholls1981}%
  \BibitemOpen
  \bibfield  {author} {\bibinfo {author} {\bibfnamefont {R.~W.}\ \bibnamefont
  {Nicholls}},\ }\href {\doibase 10.1063/1.441065} {\bibfield  {journal}
  {\bibinfo  {journal} {J .Chem. Phys.}\ }\textbf {\bibinfo {volume} {74}},\
  \bibinfo {pages} {6980} (\bibinfo {year} {1981})}\BibitemShut {NoStop}%
\bibitem [{\citenamefont {Edvinsson}\ and\ \citenamefont
  {Lagerqvist}(1985)}]{Edvinsson1985}%
  \BibitemOpen
  \bibfield  {author} {\bibinfo {author} {\bibfnamefont {G.}~\bibnamefont
  {Edvinsson}}\ and\ \bibinfo {author} {\bibfnamefont {A.}~\bibnamefont
  {Lagerqvist}},\ }\href {\doibase 10.1016/0022-2852(85)90123-7} {\bibfield
  {journal} {\bibinfo  {journal} {J. Mol. Spectrosc.}\ }\textbf {\bibinfo
  {volume} {113}},\ \bibinfo {pages} {93} (\bibinfo {year} {1985})}\BibitemShut
  {NoStop}%
\end{thebibliography}%

\end{document}